%% file: main.tex
\documentclass{article}
\PassOptionsToPackage{sort&compress}{natbib}
\usepackage{te_preprint}

\setcitestyle{numbers,super,open={},close={},citesep={,}}
\makeatletter
\AtBeginDocument{%
  \let\wsp@citep\citep
  \renewcommand{\citep}[1]{\unskip\def\wsp@keys{#1}\futurelet\wsp@next\wsp@citecheck}}
\newcommand{\wsp@citecheck}{%
  \ifx\wsp@next.\let\wsp@action\wsp@citeover
  \else\ifx\wsp@next,\let\wsp@action\wsp@citeover
  \else\let\wsp@action\wsp@citeplain\fi\fi
  \wsp@action}
\newcommand{\wsp@citeplain}{{\sffamily\wsp@citep{\wsp@keys}}}
\newcommand{\wsp@citeover}[1]{\leavevmode\rlap{#1}{\sffamily\wsp@citep{\wsp@keys}}\spacefactor\sfcode`#1\relax}
\makeatother

\usepackage[utf8]{inputenc}
\usepackage{hyperref}
\hypersetup{pdftitle={Interpretable weight-space programming builds transparent transformers and brain encoding models},
  pdfauthor={Yash Mehta, Chandan Singh, Weijia Xu, Richard J. Antonello, Adeen Flinker, Nima Mesgarani, Michael F. Bonner, Nebojsa Jojic, Jianfeng Gao}}
\usepackage{url}
\usepackage{booktabs}
\usepackage{amsfonts}
\usepackage{amsmath}
\usepackage{nicefrac}
\usepackage{microtype}
\usepackage{xcolor}
\usepackage{graphicx}
\usepackage{placeins}
\usepackage{longtable}
\usepackage{fontawesome}
\usepackage[capitalize]{cleveref}
\crefname{section}{Sec.}{Secs.}
\Crefname{section}{Sec.}{Secs.}
\crefname{table}{Table}{Tables}
\Crefname{table}{Table}{Tables}
\crefname{figure}{Fig.}{Figs.}
\Crefname{figure}{Fig.}{Figs.}
\graphicspath{{figures/}}

\newsavebox{\widefigbox}
\newenvironment{widefig}{\begin{lrbox}{\widefigbox}\begin{minipage}{\textwidth}}{\end{minipage}\end{lrbox}\makebox[\linewidth][c]{\usebox{\widefigbox}}}
\newcommand{\wideincludegraphics}[2][]{\includegraphics[width=\linewidth,#1]{#2}}

\input{tables/vision_numbers}

\newcommand{\afficon}[2][0 0 0 0]{\raisebox{-.14em}{\includegraphics[height=.95em,trim=#1,clip]{#2}}\,}
\makeatletter
\renewcommand\section{\@startsection{section}{1}{\z@}%
  {-2.4ex \@plus -.6ex \@minus -.2ex}{1.1ex \@plus .2ex}%
  {\normalfont\bfseries\fontsize{12}{15}\selectfont\color{tpink}\raggedright}}
\renewcommand\subsection{\@startsection{subsection}{2}{\z@}%
  {-1.9ex \@plus -.5ex \@minus -.2ex}{.7ex \@plus .2ex}%
  {\normalfont\bfseries\fontsize{10.5}{13}\selectfont\color{tpink}\raggedright}}
\makeatother

\usepackage{etoolbox}
\makeatletter
\def\fps@figure{t}
\def\fps@table{t}
\makeatother
\patchcmd{\abstract}{{\bfseries\color{tpaccent}Abstract\par}}{{\centering\bfseries\color{tpaccent}Abstract\par}}{}{\PackageError{main}{abstract heading patch failed}{}}

\title{\centering Interpretable weight-space programming builds transparent transformers and brain encoding models}

\author{%
\authorname{Yash Mehta}{1,2}\authorsep
\authorname{Chandan Singh}{1}\authorsep
\authorname{Weijia Xu}{1}\authorsep
\authorname{Richard J. Antonello}{3}\authorsep
\authorname{Adeen Flinker}{4}\authorsep
\authorname{Nima Mesgarani}{3}\authorsep
\authorname{Michael F. Bonner}{2}\authorsep
\authorname{Nebojsa Jojic}{1}\authorsep
\authorname{Jianfeng Gao}{1}}
\affiliations{\textsuperscript{1}\afficon[0 0 265.6pt 0]{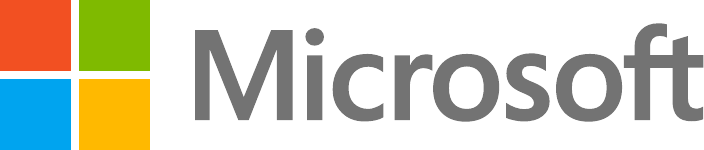}Microsoft Research\qquad
  \textsuperscript{2}\afficon[0 0 236.4pt 0]{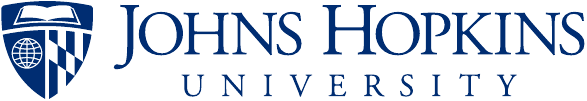}Johns Hopkins University\qquad
  \textsuperscript{3}\afficon[.5pt 18.5pt 241.8pt .7pt]{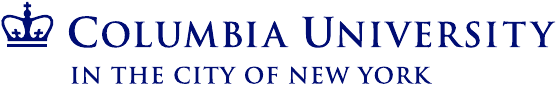}Columbia University\qquad
  \textsuperscript{4}\afficon{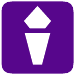}New York University}
\renewcommand{\weblink}[2]{{\color{tpink}#1} & \href{https://#2}{\nolinkurl{#2}}}
\links{\weblink{\faGithub}{github.com/csinva/weight-space-programming-autoresearch}}

\begin{document}

\maketitle

\begin{abstract}
Neural networks are powerful predictors, yet their opacity limits their usefulness in science, where the goal is explanation rather than mere prediction. In neuroscience, for example, deep encoding models accurately predict brain activity in response to natural stimuli, but struggle to reveal which stimulus features drive those predictions. Here we introduce interpretable weight-space programming (iWSP), a method in which an LLM agent iteratively builds interpretable models by explicitly programming their weights and features in code, without gradient-based training (backpropagation). Without backpropagation to tune parameters, the agent must draw on its prior knowledge and reason over previous trials to hypothesize, ``hand-craft'', and refine human-interpretable computations mirroring the workflow of a human scientist.
LLM audits confirm the interpretability of the hand-crafted models. We first validate iWSP by tasking it with constructing transformers for canonical benchmarks such as arithmetic and sentiment classification, finding that LLM agents can successfully craft compact, accurate networks directly. Next, we apply iWSP to build neural encoding models across vision, language, and audition. In the visual cortex, iWSP outpredicts existing interpretable baselines, uncovers distinct computations across early/higher visual areas, and generalizes to unseen data. In language fMRI, iWSP surfaces a confound in the data and in auditory ECoG, it systematically discovers interpretable acoustic features that predict neural activity. Together, these results establish iWSP as a general paradigm for aligning high predictive performance with human-interpretable scientific discovery.

\end{abstract}

\section{Introduction}

\begin{figure}[t]
\begin{widefig}
  \centering
  \wideincludegraphics{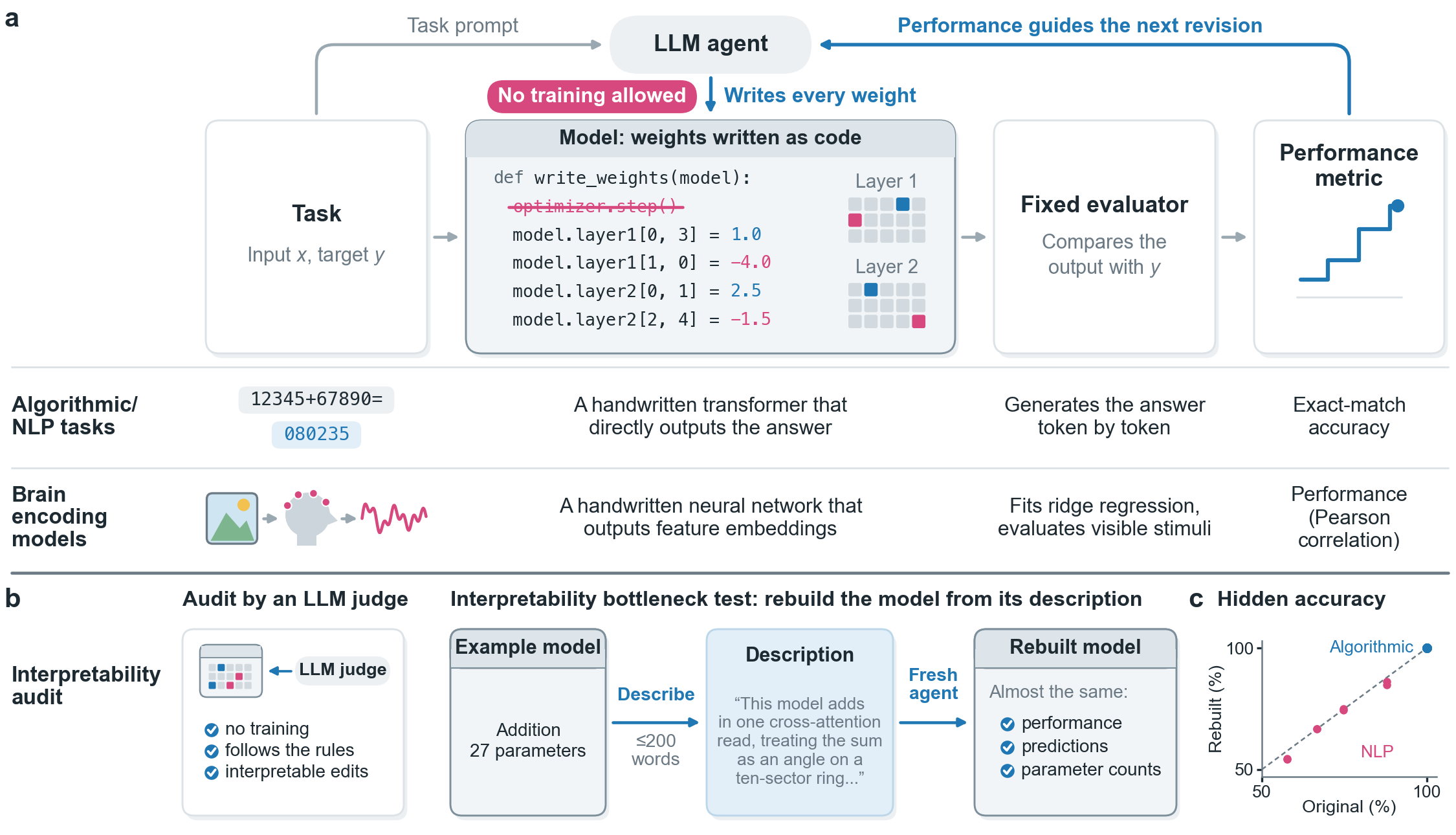}
  \caption{\textbf{The iWSP loop and its interpretability audit.} \textbf{a}, An LLM agent reads a task prompt and writes a model whose weights are set directly in code. No training is allowed.
  A fixed evaluator compares the model's output with the target $y$ and returns a performance metric, which the agent uses to revise the code.
  On algorithmic/NLP tasks the model is a neural network whose weights are all set by the agent to generate the exact answer.
  For brain encoding, the hand-written network computes fixed features, and the evaluator fits a ridge regression from those features to recorded responses and evaluates it on visible stimuli.
  \textbf{b}, After the search, an LLM judge audits every saved model for training, rule violations and uninterpretable edits. To test interpretability, an agent (Opus 5.5) describes the best model in at most 200 words and a fresh session of the agent rebuilds it from that text alone, shown here for an \textit{addition} task. \textbf{c}, In the algorithmic/NLP tasks for example, we find that the original and rebuilt models achieve very similar performance (2 rebuilds per task).}
  \label{fig:loop}
\end{widefig}
\end{figure}

In the early days of artificial neural networks, McCulloch and Pitts wrote every connection of small networks by hand to make them compute logical functions\citep{mcculloch1943logical}.
The values of these weights were interpretable, allowing one to understand how the models worked.
In the modern era, as massive networks trained with backpropagation have improved their performance, models have sacrificed this interpretability.
The resulting opacity raises concerns around regulatory compliance, safety, and alignment\citep{goodman2017european,amodei2016concrete,gabriel2020artificial,singh2024rethinking}.
It is particularly limiting in fields such as neuroscience\citep{jain2024computational} and social science\citep{ziems2024can}, where trustworthy interpretation is often itself the end goal.
Neuroscience has seen the rise of large neural-network based ``encoding models'' that predict brain responses to natural stimuli, like images or text\citep{naselaris2011encoding,yamins2014performance,cadena2019deep,zhuang2021unsupervised,konkle2022self,jain2018incorporating}.
However, these models are often difficult to interpret, making it difficult to connect features of the input stimulus to brain activity\citep{toneva2019interpreting,goldstein2022shared,caucheteux2023evidence,antonello2023scaling}.

To tackle this challenge, we introduce \emph{interpretable weight-space programming} (iWSP).
iWSP guides a frontier agent to directly write a model while disallowing any training.
Then, a fixed evaluator measures its performance, and the agent revises it iteratively (\cref{fig:loop}).
Disallowing training requires the agent to reason and implement each change in text/code, so weight changes are generally guided by the agent's prior knowledge and each change is accompanied by a written explanation and resulting change in performance.

iWSP builds on a vast literature.
One line of work builds inherently interpretable models\citep{friedman2023learning,rudin2019stop,huth2016natural,liu2024infini,singh2023augmenting,koh2020concept,kim2024interpretable,sun2024concept,team2026scaling,benara2024crafting,feng2026human,ludan2023interpretable,tan2024interpreting}, though their constrained form often limits accuracy on hard problems.
To get around this limit, recent work uses autoresearch to discover effective programs to solve hard problems\citep{romera2023mathematical,novikov2025alphaevolve,georgiev2025mathematical,singh2026agentic,kasenberg2026ai,lu2024ai,gottweis2025towards,swanson2025virtual,ifargan2024autonomous,tang2025ai,mitchener2025kosmos}.
Such programs can be translated into neural networks using a domain-specific language\citep{weiss2021thinking,lindner2023tracr} or alternatively these weights can be generated by training a network to do so\citep{ha2016hypernetworks,chauhan2024brief,stanley2009hypercube,schurholt2022hyper,peebles2022learning,knyazev2021parameter,han2026survey}.
In weight-space programming\citep{takehi2026weightspace}, the agent instead reads and edits the weights itself, potentially guided by mechanistic interpretability approaches, such as computing activation differences\citep{elhage2021mathematical,wang2022interpretability,turner2023steering,li2023inference,meng2022locating}.
We extend this work with a simple modification:
we require that weight-space edits be interpretable and then iteratively update weights to optimize a performance objective.
To ensure interpretability, we audit the resulting models (both manually and with LLMs as a judge), and quantitatively test whether they can be rebuilt from a short text description (\cref{fig:loop}b).

We first evaluate iWSP on relatively simple algorithmic/NLP tasks.
Using frontier agents (e.g. Claude Opus 5.5), iWSP writes transformers from scratch that effectively solve 13 different algorithmic tasks, and achieve respectable performance on 3 out of 4 standard NLP tasks like sentiment classification (\cref{sec:wsp}).
We then study iWSP in three settings that predict brain responses to naturalistic stimuli.
First, in visual cortex, an iWSP model written on one functional magnetic resonance imaging (fMRI) dataset predicts responses in another better than interpretable baselines, and its computations separate early and higher visual areas (\cref{sec:vision}).
Second, in language fMRI, the interpretability of iWSP is useful for debugging, helping to identify a dataset confound~(\cref{sec:story}).
Third, in electrocorticography (ECoG) responses to audio (\cref{sec:ecog}), iWSP finds accurate interpretable models that can improve the predictive performance of black-box models.
Taken together, these results suggest that frontier agents can help close the gap between highly accurate neural networks, and ones that we can usefully understand.

\section{Method: Interpretable weight-space programming}
\label{sec:method}

Motivated by the search for interpretable models that can help scientific domains, we seek to improve the performance of human-understandable neural networks.
To do so, we follow an approach similar to weight-space programming\citep{takehi2026weightspace}, which prompts an agent to directly read and edit neural network weights.
We study \textit{interpretable weight-space programming} (iWSP), where we further 
require that weight-space edits be interpretable and then iteratively update weights to optimize a performance objective.
Interpretability is suggested in the prompt, and then enforced by an auditing agent that judges whether model changes at each iteration are human-interpretable.
This approach enables an agent to search for an interpretable model much like a human would, by using its priors and domain knowledge.

At a high level, each iWSP run has the same setup (\cref{fig:loop}).
A frontier coding agent is given a prompt that describes the task, and a harness that includes a starting file and a non-editable evaluation script that evaluates the current model.
The prompt and harness forbid training, including optimizers, fitting loops, and downloading pretrained weights.
The agent remains free to use known intepretable operations (e.g. Gabor filters), but must implement them using only its given basic packages, such as pytorch.
It then repeatedly writes a model, evaluates it on a ``visible set'' of examples and revises it for a fixed number of iterations.
Throughout, the agent is prompted to search for a human-understandable solution.

\paragraph{Interpretability audits} Each saved model (and the agent's corresponding logs) is audited by a separate agent to test for the interpretability of the edits and any violations of the constraints outlined in the prompt and harness (\cref{fig:loop}b).
As an additional check on whether a model uses any training data explicitly, we verify that rerunning each model's weight-writing code on altered task data leaves its weights unchanged (algorithmic and NLP tasks).
After the run concludes, the models are reevaluated on a ``hidden set'' that they never saw during the iterative loop.
We further test the interpretability of the final models by checking whether the models can be accurately described with a short text description using a frontier LLM (Opus 5.5).
We ask the LLM to describe the best model for each task in at most 200 words, and then give only that description and the task to a fresh session, which rebuilds the model (\cref{fig:loop}b).
We then evaluate how well the rebuilt model reflects the original, e.g. does it maintain similar predictive performance?

\paragraph{Algorithmic/NLP tasks.}
In the algorithmic/NLP tasks, each run starts from one Python file and one instruction document. The file defines a small GPT-style transformer, with three layers of width 64 by default, a function \texttt{write\_weights(model, task)} for the agent to fill in, and fixed code that evaluates the result. The prompt names the task and tells the agent to follow the document until it has run 30 to 50 iterations, depending on the task. The document states the rules, including:

\begin{quote}
\small
\textit{The model is NEVER trained. The agent must write the weights directly into} \texttt{interpretable\_transformer.py}\textit{.} [\ldots] \textit{What you CANNOT do: Run any kind of training, gradient descent, optimizer step, fitting loop, or backprop. The model parameters must be set in closed form.} [\ldots] \textit{The goal is simple: maximize} \texttt{accuracy} [\ldots] \textit{All else being equal, more interpretable is better. A small improvement that adds ugly complexity is not worth it. Number of model parameters can help serve as a proxy for interpretability, but the real criterion is human-understandable structure in the weights.}
\end{quote}

The agent may change the architecture and hard-code constants, including hand-built attention circuits and lookup tables; it may not compute the answer with anything outside the network (e.g. a Python function). Each run has its own folder, and the agent works in a restricted Claude Code session with six file and shell tools, no web access and file reads confined to that folder.
Its only shell command, apart from read-only listings, runs this file. The harness cannot stop the agent's own Python from training or from computing the answer outside the network, so those rules remain instructions that we audit afterward.
The agent learns the task format by printing examples from inside its code, and after each iteration it sees its visible accuracy on a fixed visible set of 200 examples, where the answer is generated token by token and must match exactly.
For these experiments, the iWSP agent is Claude Code running Opus 5.5, Sonnet 5.5 or Haiku 5.5 and audits for constraint-following are done by Opus 5.5.

\paragraph{Brain encoding models.}
In our experiments involving brain data, the agent writes a model $f$ that maps each stimulus $x$ to features $f(x)$.
A fixed evaluator then fits a ridge regression model $W$ from $f(x)$ that linearly predicts recorded responses $y$ on training stimuli, predicts $\hat{y}=Wf(x)$ on the visible set, and returns the pearson correlation between $\hat{y}$ matches $y$. The agent receives only this metric and every candidate within a dataset uses the same stimulus split, ridge regression procedure, and metric.
In addition to the standard constraints above, the rules forbid a model from reading any of the brain responses and the evaluation code rejects a candidate with 10{,}000 or more features. Unless otherwise specified, the agentic system is Claude Code with Opus 5.5; the visual cortex search uses \visSearchModel{} (\cref{sec:vision}). The visual cortex search runs in a sandbox with no network and no access to the datasets, run directories or repository. The agents work in their own sandbox, which shows only their workspace, and they receive performance metrics through a restricted interface.
The story-listening fMRI and ECoG searches do not isolate in a sandbox and instead rely on the prompt with a post-hoc audit (\cref{sec:story,sec:ecog}).

In total, the paper's experiments cost about \$17{,}500 at API list prices, including estimated exploratory runs (\cref{tab:costs}).

\section{Algorithmic/NLP results: iWSP builds effective, interpretable transformers}
\label{sec:wsp}

\paragraph{Experimental setup.}
To build our intuition, we first evaluate iWSP on 17 diverse tasks.
The tasks include 13 algorithmic tasks such as multiplication, sorting, edit distance and date normalization (inspired by the AdderBoard addition task\citep{papailiopoulos2026adderboard}), and four NLP tasks: 6-class question classification on TREC\citep{li2002learning}, binary sentiment classification on SST-2\citep{socher2013recursive}, binary paraphrase detection on MRPC\citep{dolan2004unsupervised} and three-way natural language inference on SNLI\citep{bowman2015large} (see full descriptions in \cref{tab:tasks}).
We run iWSP on each task twice with each of Claude Opus 5.5, Sonnet 5.5 and Haiku 5.5, for about 1{,}340 iterations per LLM (about 39 per run). The hidden set is 2{,}000 new examples for algorithmic tasks and the official test split for NLP tasks. After the runs, we audit the iWSP models as described in \cref{sec:method} and exclude the few that do not pass from the results (see full details in \Cref{tab:wsp_models}).

\begin{figure}[t]
\begin{widefig}
  \centering
  \wideincludegraphics{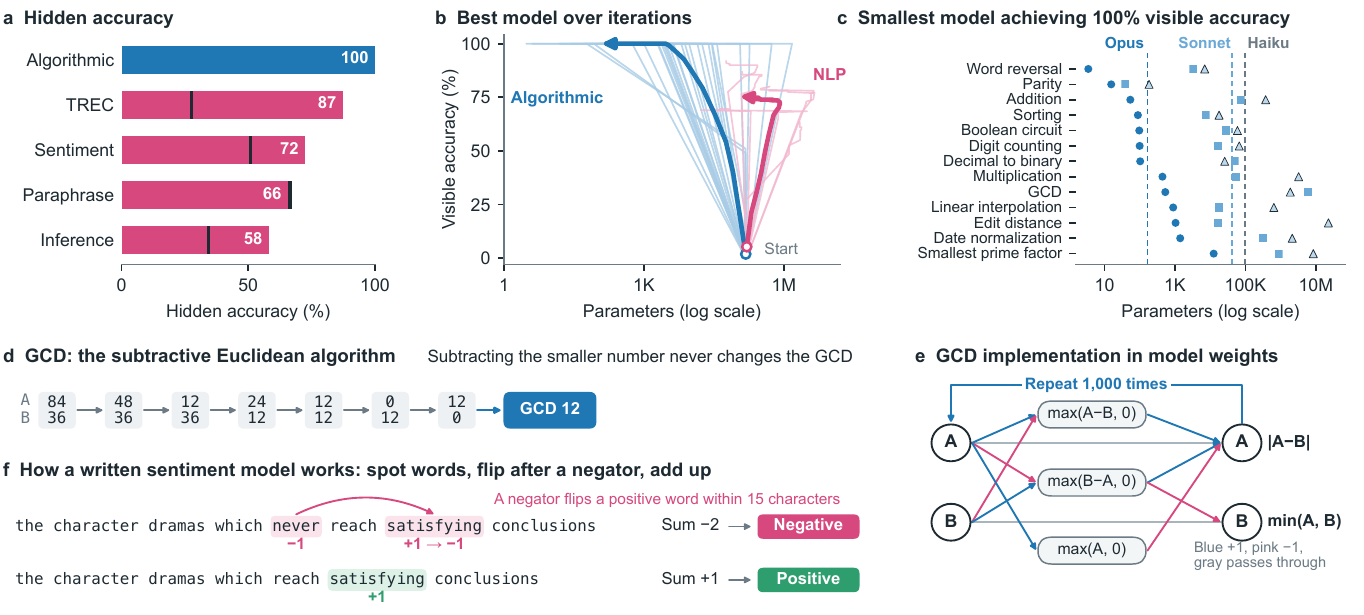}
  \caption{\textbf{iWSP writes exact, readable models for algorithmic tasks and weaker ones for NLP tasks.}
  Results in this figure use Claude Opus 5.5 unless noted.
  \textbf{a}, Models reach 100\% hidden accuracy on algorithmic tasks but not on NLP tasks, where paraphrase stays below its baseline. Bars show hidden accuracy over 2 runs per task, and black lines mark always giving the most common answer.
  \textbf{b}, The searches both improve over iterations: algorithmic runs reach 100\% visible accuracy early and then find more compact models, while NLP runs taper off. Thin lines follow each run's best model so far, and thick lines average each group from the open circle to the arrowhead.
  \textbf{c}, The ability to find small accurate models depends on the LLM: Opus 5.5 finds far smaller models that reach 100\% visible accuracy than Sonnet 5.5 or Haiku 5.5. Markers show geometric means over 2 runs and dashed lines show medians.
  \textbf{d}, For the algorithmic tasks, written models often translate a known algorithm into a transformer. For finding the greatest common divisor (GCD), the model uses the subtractive Euclidean algorithm, shown here on one input pair.
  \textbf{e}, iWSP implements this algorithm using a simple repeating block of ReLU units.
  \textbf{f}, For NLP tasks, iWSP builds crude interpretable tools to represent the intricacies of language, such as a rule-based negator for sentiment classification.}
  \label{fig:params}
\end{widefig}
\end{figure}

\paragraph{Models generalize on algorithmic tasks and on 3 out of 4 NLP tasks.}
We first describe the Opus runs. All 26 runs on algorithmic tasks reach 100\% accuracy both on the visible set and the hidden set (\cref{fig:params}a), often by implementing a known algorithm in the weights.
The NLP tasks are harder as they have no known full solution.
Nevertheless, on 3 of the 4 NLP datasets iWSP builds models that achieve respectable performance well above a naive majority baseline (it fails on the relatively difficult Paraphrase task).

\paragraph{iWSP compresses models after making them accurate.}
Over iterations, both types of tasks improve in their accuracy and then start to compress the model parameters (\cref{fig:params}b). On average, algorithmic runs reach 99\% visible accuracy within the first tenth of their iterations and spend the rest compressing from about 3{,}400 parameters to about 200. The NLP runs' performance plateaus after about 75\% of their iterations and stay near 190{,}000 parameters.
On algorithmic tasks, the smallest model that reaches 100\% visible accuracy has a median of 166 parameters (\cref{fig:params}c), and in the median run it is 29 times smaller than the first model to reach 100\%.

\paragraph{Bigger LLMs mostly achieve similar performance but are better at compressing models.}
When comparing different Claude models (Opus 5.5, Sonnet 5.5, and Haiku 5.5), we find that they generally all achieve about 100\% accuracy on the algorithmic tasks, but that Opus and Sonnet outperform Haiku on the NLP tasks (\cref{tab:wsp_models}).
Beyond predictive accuracy, better LLMs are better able to compress an accurate model (\cref{fig:params}c), with the smallest model that reaches 100\% visible accuracy having a median of 166 parameters for Opus, 41{,}600 for Sonnet and 100{,}000 for Haiku.

\paragraph{iWSP models translate known algorithms into transformers.}
For algorithmic problems, the agent tends to implement known algorithms rather than conceiving of new ideas.
For example, the \textit{edit distance} task uses one block per row of the dynamic-programming table, and \textit{multiplication} adds shifted partial products.
For \textit{greatest common divisor} (GCD), the subtractive Euclidean algorithm\citep{yao1975analysis} uses the fact that any common divisor of $A$ and $B$ also divides $A-B$, so replacing the pair with $(|A-B|, \min(A,B))$ preserves the greatest common divisor until the pair reaches $(g, 0)$ (\cref{fig:params}d).
iWSP implements this as a one block of three ReLU units, repeated 1{,}000 times (\cref{fig:params}e).
Note that 100\% performance on our eval tasks does not guarantee correct outputs on all inputs, there could be adversarial inputs not considered in the task or generalizations for which the network completely fails, e.g. our GCD test only specifies 3-digit inputs.

\paragraph{iWSP NLP models are crude but interpretable.}
Without a known algorithm, iWSP builds simple rules that approximate what an NLP task requires.
For example, the best sentiment classification model has three stages, each built from standard transformer parts (\cref{fig:params}f). A shift register made of attention heads feeds one neuron for each of 1{,}515 hand-written word patterns, a second attention layer looks back for a negator and flips positive evidence that follows it, and a final layer adds up the evidence.
The negator is a crude rule, but is effective at composing terms in a way that simple words/ngrams fail to.

\paragraph{Vetting interpretability: a 200-word description is sufficient to rebuild a model.}
Beyond qualitative analysis, we further run the text bottleneck check described in \cref{sec:method} (\cref{fig:loop}b).
We find that all 13 rebuilt models (with 2 seeds each) reach 100\% hidden accuracy on the algorithmic tasks, and on NLP tasks rebuilt models lose about 1.4 points of hidden accuracy on average (\cref{fig:loop}c). We further audit that this is done properly: (i) no rebuilt model computes its weights from task data (either through training or corpus statistics),
(ii) rerunning the weight-writing code of all 34 rebuilt models on altered task data leaves their weights unchanged,
and (iii) no session read files outside its sandbox.
The rebuilt models are also much closer to the original than a control that rebuilds models without a description.
On algorithmic tasks, rebuilt models have 1.74 times the original's parameters on a geometric mean, against 2{,}043 times for the control (\cref{fig:wsp_bottleneck}b).
On NLP tasks, rebuilt models also perform closer to the original than the control does, e.g. their hidden-set predictions correlate significantly more on average (0.74 against 0.51; \cref{fig:wsp_bottleneck}c).

\section{Neuroscience results: using iWSP to build brain encoding models}
\label{sec:results}

\subsection{iWSP builds highly predictive, interpretable models of the visual cortex}
\label{sec:vision}

\begin{figure}[!t]
\begin{widefig}
  \centering
  \wideincludegraphics{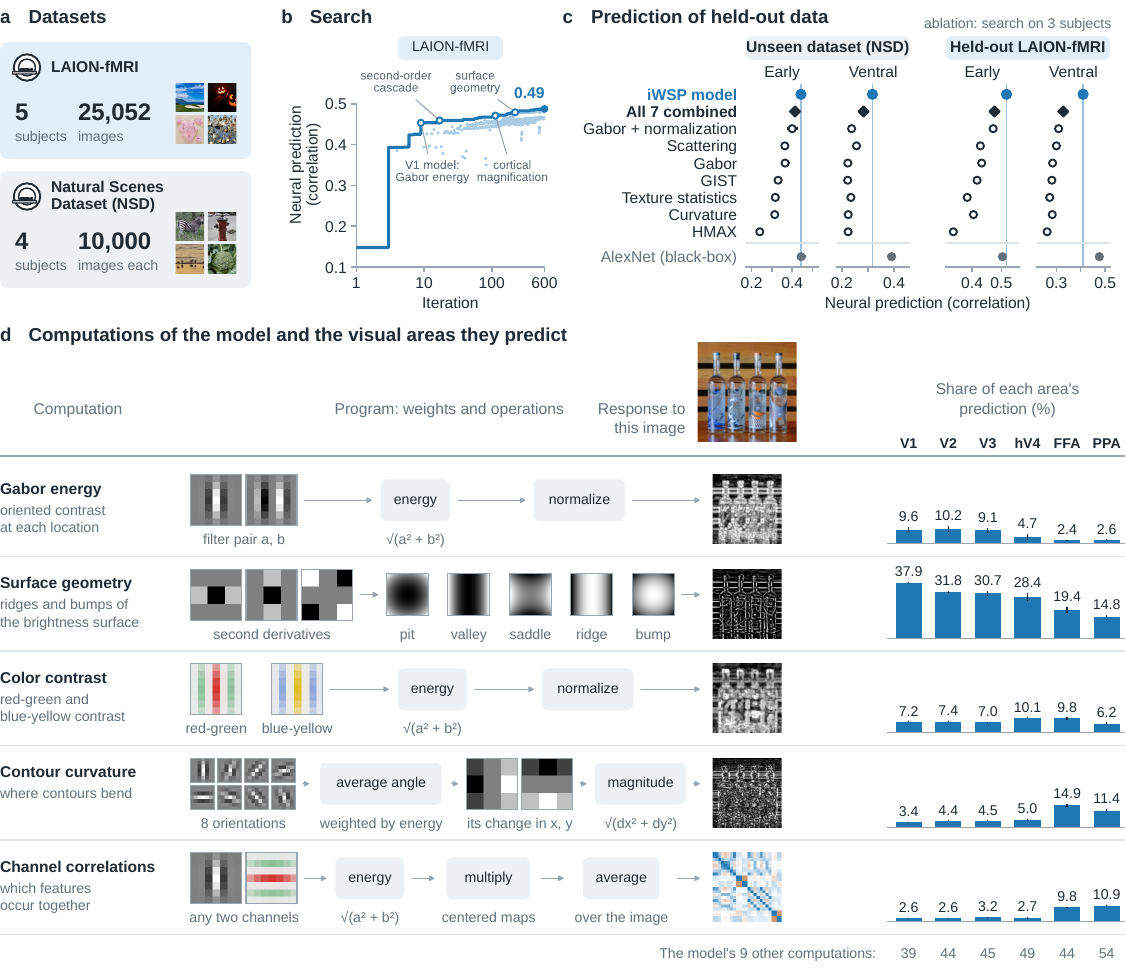}
  \caption{\textbf{iWSP writes an interpretable visual encoding model that generalizes across subjects and datasets.}
  \textbf{a}, The two fMRI datasets, with example stimuli.
  \textbf{b}, We run iWSP on five subjects using the data from LAION-fMRI and find that prediction performance (averaged over the early and ventral visual streams) improves over iterations.
  \textbf{c}, We then evaluate the final iWSP model on the held-out NSD dataset, and find that it outperforms interpretable baselines and slightly underperforms a black-box AlexNet baseline.
  Additionally, to test held-out performance on LAION fMRI, we rerun iWSP using only 3 subjects and test on the held-out 2 subjects and held-out images; we find the same performance patterns.
  \textbf{d}, We interpret the best model from the search in \textbf{b}, which consists of 14 computations (see \cref{fig:fmri_attribution}).
  We highlight five of the computations here, showing  the filters and operations of one computation, its response to the example image (for channel correlations, the matrix of correlations between channel maps), and the share of each visual area's prediction that it accounts for.
  Error bars in \textbf{c} and \textbf{d} show the standard error of the mean over subjects, which is often within the points}
  \label{fig:fmri}
\end{widefig}
\end{figure}

In visual neuroscience, the most accurate encoding models are deep networks trained for object recognition, self-supervised learning, or image--text alignment\citep{yamins2014performance,khaligh2014deep,cadena2019deep,zhuang2021unsupervised,konkle2022self,schrimpf2018brain,chen2025universal}, whose learned weights make it hard to state what image features they compute.
In contrast, classical models built from explicit operations such as Gabor filters\citep{kay2008identifying}, divisive normalization\citep{heeger1992normalization}, scene gist\citep{oliva2001modeling}, or texture statistics\citep{portilla2000parametric} are interpretable but predict brain responses less accurately. Here we use iWSP to bridge this gap by searching for a neural network that builds interpretable features from images that are highly predictive of brain responses (\cref{sec:method}).

\paragraph{Experimental setup.}
We run iWSP on LAION-fMRI\citep{zerbe2026laion} (7\,T visual cortex responses from five subjects viewing 25{,}052 natural images) and hold out the Natural Scenes Dataset (NSD) for out-of-dataset testing (\cref{fig:fmri}a). Following \cref{sec:method}, each candidate maps raw images to fewer than 10{,}000 features using fewer than 100{,}000 explicit parameters, and a fixed ridge regression fit on each subject's unique images is scored by Pearson correlation on shared validation images. Search maximizes mean correlation pooled across subjects and visual areas while also returning per-subject and per-area scores (such as V1, V2, V3, hV4, FFA, and PPA). Once search on all five LAION-fMRI subjects finishes, we freeze the best program (the iWSP model) and test it zero-shot on NSD. To test unseen subjects and images within LAION-fMRI, we also repeat the search on \visSearchSubjects{} subjects and half of the shared images, scoring its best model once on the other \visFinalSubjects{} subjects and \visLaionImages{} shared images. On both datasets, we compare iWSP against seven classical interpretable baselines (including Gabor filters with divisive normalization and Portilla--Simoncelli texture statistics\citep{portilla2000parametric}), a single ridge regression fit to all seven combined, and pretrained AlexNet\citep{krizhevsky2012imagenet} (\cref{sec:app_visual_model}).

\paragraph{iWSP builds from classical V1 filtering to 14 explicit computations.}
Across the \visCandidates{} search iterations on LAION-fMRI, performance steadily increases  (\cref{fig:fmri}b), with most of the gain occuring within the first \visNinetyIterations{} iterations as the agents recover a classical V1 front end of multi-scale oriented Gabor filters and divisive normalization\citep{heeger1992normalization} on luminance and color-opponent channels. The final \visModelLines{}-line program produces 9{,}996 features from 14 explicit computations (\cref{sec:app_visual_model}), five of which appear in \cref{fig:fmri}d:
(1) \emph{Gabor energy} and (2) \emph{color contrast} compute locally normalized filter energy on luminance and red--green or blue--yellow channels;
(3) \emph{surface geometry} uses second derivatives of brightness to classify local shading into pits, valleys, saddles, ridges, and bumps; (4) \emph{contour curvature} tracks how fast local orientation changes across space; and (5) \emph{channel correlations} average pairwise products of feature maps to capture co-occurrence. Confirming that the program is readable, fresh sessions given only a 185-word summary rebuild models that reach 0.451 pooled correlation (versus 0.488) and match the original's predictions more closely than AlexNet conv2, while rebuilds across eight search checkpoints retain 96\% of original performance on average (\cref{fig:fmri_bottleneck}; \cref{sec:app_vision_bottleneck}).

\paragraph{iWSP generalizes to an unseen dataset and to held-out subjects.}
Evaluating the same iWSP hand-crafted features on NSD (\cref{fig:fmri}c, left) shows that they transfer across datasets and subjects. In every one of the \visNsdSubjects{} NSD subjects, iWSP beats all 7 interpretable baselines in both early (\visNsdEarly{} versus \visNsdEarlyBest{} for the best baseline) and ventral visual cortex (\visNsdVentral{} versus \visNsdVentralBest{}), and outperforms a single ridge regression fit jointly to all seven baselines (\visNsdVentralCombined{} in ventral visual cortex). Compared with pretrained AlexNet, iWSP matches black-box accuracy in early visual cortex while trailing in ventral visual cortex (\visNsdAlexVentral{}). Repeating the search on \visSearchSubjects{} LAION-fMRI subjects (\cref{fig:fmri}c, right) replicates every comparison on the \visFinalSubjects{} held-out subjects and hidden images (early \visLaionEarly{} versus \visLaionEarlyBest{}; ventral \visLaionVentral{} versus \visLaionVentralBest{}), with identical accuracy on hidden and visible images in the visible subjects (\cref{sec:app_visual_model}).

\paragraph{Different computations explain early versus category-selective visual areas.}
Because the ridge regression sums over features, each computation's covariance with the measured response gives its exact percentage share of a voxel's test correlation, summing to 100\% in every area (\cref{sec:app_visual_model}). Although search optimizes a single pooled score without anatomical labels, these shares shift systematically across the visual hierarchy (\cref{fig:fmri}d and \cref{fig:fmri_attribution}). Early areas rely more on local shading and edge energy: local-mean surface geometry falls from 38\% in V1 to 15\% in PPA, and local-mean Gabor energy drops from about 10\% in V1--V3 to under 3\% in FFA and PPA, showing that the final V1 ridge regression relies more on later-added surface geometry than on the initial Gabor stage. Category-selective areas instead draw more on mid-level structure, with contour curvature rising from 5\% or less in V1--hV4 to 15\% in FFA, channel correlations rising from about 3\% in V1--hV4 to about 10\% in FFA and PPA, and color contrast peaking in hV4 and FFA. Because correlated computations share 88\% to 96\% of the model's squared correlation, we verify these profiles by refitting the ridge regression without each computation (\cref{tab:fmri_unique}): in all five subjects, local-mean Gabor energy uniquely explains 1.4\% to 2.0\% in V1--V3 and none in FFA and PPA, local-mean surface geometry uniquely explains 0.7\% to 1.5\% in every area, and channel correlations uniquely explain 4.9\% in FFA and 1.2\% in PPA versus none in V1--V3.

\subsection{iWSP identifies a data confound in story-listening fMRI}
\label{sec:story}

We next apply iWSP to a language fMRI dataset, in which subjects listen to narrative stories and the agent builds a model to predict responses from the words.
iWSP rewards anything that raises its prediction performance, including confounds.
Because iWSP models are interpretable, we can use this as an advantage: by reading the best-performing iWSP model, we identify a data confound at the edge of each fMRI scan that, once understood, can be removed.

\paragraph{Experimental setup.}
Encoding models of fMRI responses to natural language are often used to study language processing in the brain\citep{huth2016natural,schrimpf2021neural,antonello2023scaling,jain2024computational,singh2025evaluating,mitchell2008predicting,wehbe2014simultaneously,nishida2018decoding,kumar2022shared,caucheteux2023evidence}. We use data from two recent studies\citep{lebel2022natural,tang2023semantic}, with fMRI responses from three subjects listening to more than 20 hours of narrative stories from podcasts. We analyze one subject (UTS03) across 84 story scans, each 157 to 521 volumes long at one volume per 2\,s (one TR). Each scan covers 95{,}556 voxels. The preprocessed release z-scores every voxel within each scan and has already removed 20 TRs per scan. The search fits ridge regression on 8 stories and evaluates it on 3 other stories, which form the visible evaluation set: the agent sees this performance at every iteration. After the search, we also test selected models on a hidden evaluation set of 10 other stories, drawn at random and never seen by the agent, with ridge regression still fit on the original 8 stories. Unless noted, performance below is visible performance.

We run iWSP with Claude Opus 4.7, and use a fixed evaluator that (i) aligns the features to the TRs, (ii) adds delays of one to four TRs for the slow hemodynamic response, (iii) fits ridge regression and (iv) returns the mean correlation across voxels on the visible stories. For every word in a story, the model is fed the word and the 10 words before it, and returns one feature vector per word, which the evaluator then aligns to the TRs. As a black-box baseline, we feed these same contexts to GPT-2 XL\citep{radford2019language} and take the hidden state of the final token at layer 24 out of 48 (chosen via cross-validation over the layers 12, 24, and 36).

\begin{figure}[t]
\begin{widefig}
  \centering
  \wideincludegraphics{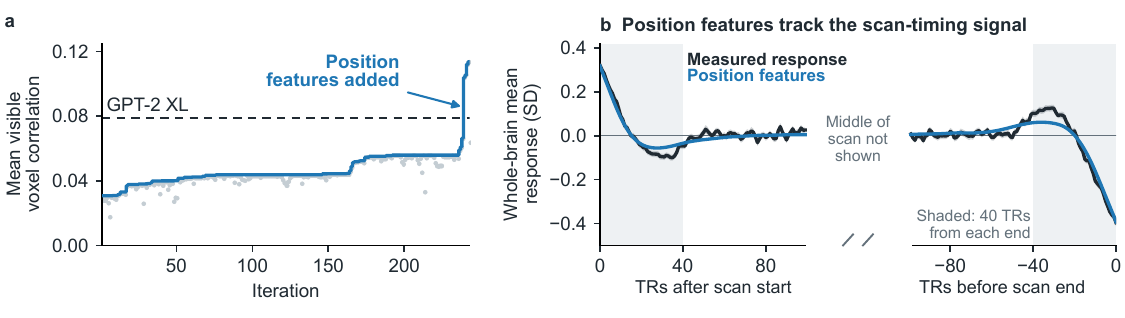}
  \caption{\textbf{A jump in performance during iWSP iterations identifies a dataset confound.}
  \textbf{a}, After growing gradually, the performance jumps in a single step by adding features that code for a word's position in the story rather than its semantics.
  \textbf{b}, Debugging this feature identifies that the beginning/end of each scan has a characteristic pattern unrelated to the content of the stories (black line).
  Predicting using just position features (blue line) can track this measured response well, providing potentially misleading improvements for models that are not explaining genuine signal in brain activity.
  }
  \label{fig:story}
\end{widefig}
\end{figure}

\paragraph{A data confound unrelated to language processing.}
For most of the iWSP run, features of word content improve slowly, scoring well below the 0.079 correlation achieved by GPT-2 XL (\cref{fig:story}a).
Then, performance suddenly jumps from 0.067 to 0.104 in a single step.
We find that the new features added at this step surprisingly do not encode a word's semantic content, but rather only its position in the story. Further analysis shows that these position features alone reach about 0.090, above even the strong GPT-2 XL black-box baseline.
To investigate these position features, we average the response across voxels and scans at each position in the scan, and find that it starts abnormally high and ends abnormally low compared to the rest of the scan (\cref{fig:story}b, black line); this signal appears in all three subjects (\cref{fig:scan_edges_subjects}).
The position features closely track this signal: fit to half of the scans and tested on the other half, they follow this average response curve at a correlation of 0.97, and the jump they cause also appears on the hidden set (\cref{tab:story_hidden}).

This signal is potentially problematic because it follows scan timing rather than story content, potentially providing misleading changes in measured performance for different encoding models.
This signal is not entirely unknown: the dataset release removes 10 TRs at each end of every scan to avoid onset transients and poor detrending\citep{lebel2022natural}, and Antonello et al.\ noted that responses in some voxels drift over the first minute of each scan\citep{antonello2023scaling}, which lets long-context features inflate measured performance by up to 20\%.
To correct for this, they excluded the first 100 seconds of each test story when evaluating, a convention later adopted by other work\citep{moussa2024improving,binhuraib2025litcoder}.
Our results suggest that for future studies that are sensitive to context lengths or story position, they should go further in addressing this confound by trimming the edges of each scan (in training as well as testing) by roughly 40 TRs.
Note that this dataset remains of exceptional quality and that the many valuable conclusions drawn from it (which essentially all use word-level or short-context encoding models) stand;
identifying this confound is more preventative, as non-experts and agents continue doing analyses on this type of data in the future.

\paragraph{Interpreting the fitted iWSP model.}
After this trimming, we find that iWSP can still build interpretable models that focus on semantics, performing on par with GPT-2 XL embeddings (\cref{fig:trimmed}).
Each feature of these models is readable and has its own ridge regression weights, so it can be traced onto the cortex; \cref{fig:story_feature} shows one feature that encodes words related to mental processing or speech.
As in the previous sections, we again describe the iWSP models (from 8 points along the iWSP iterations) in a text description of at most 200 words. We find that this is sufficient for a fresh agent to rebuild each of them closely enough to achieve similar performance, including the story-position features of the first model after the jump (\cref{fig:story_bottleneck}).

\subsection{iWSP identifies interpretable, performant features for predicting ECoG responses to audio}
\label{sec:ecog}

We next study ECoG recordings of responses to natural audio, where we compare iWSP with embeddings from Whisper\citep{radford2022robust}, a standard, pretrained speech model, which is established as a highly performant feature space for encoding~\citep{antonello2023scaling}.
We find that interpretable raw-waveform features from iWSP reach similar prediction performance as Whisper, and that they can explain part of the response that Whisper fails to predict.

\paragraph{Experimental setup.}
We predict high-gamma activity at 30\,Hz from 166 intracranial electrodes in one patient, recorded continuously over 7 days\citep{zada2025podcast,goldstein2023deep}.
The stimulus is the raw audio from the room, cut into segments of about two minutes. Ridge regression is fit on 500 segments (about 16 hours of data). The visible evaluation set is 23 consecutive segments (about 44 minutes) during which a podcast was playing: the agent sees the prediction performance on this set at every iteration, and uses it to make inclusion and exclusion decisions about its feature choices. The hidden evaluation set is another 317 speech-rich segments from the same session (about 10 hours of data).

We run iWSP with Claude Opus 4.7, editing a model that maps the audio waveform to features. A fixed evaluator adds 4 time-lagged delays, shifts the features by 100\,ms, fits a ridge regression for each electrode and returns the mean correlation on the visible evaluation set. Whisper-large-v3 encoder features go through the same evaluator. As an interpretable baseline, five isolated Claude Opus 5.5 sessions each write, in one shot and without performance feedback, the feature space an expert would assemble from standard acoustic features (spectrograms, cepstral coefficients, pitch and loudness), using any open-source library. These textbook features also go through the same evaluator.
The search ran 902 iterations, plateauing after about 500 iterations.
Partway through the loop, we switched to optimizing the 125 surface-grid electrodes, where Whisper's advantage was concentrated. We still evaluate on all 166 electrodes using the hidden set.

\begin{figure}[t]
\begin{widefig}
  \centering
  \wideincludegraphics[trim=0 12 0 0,clip]{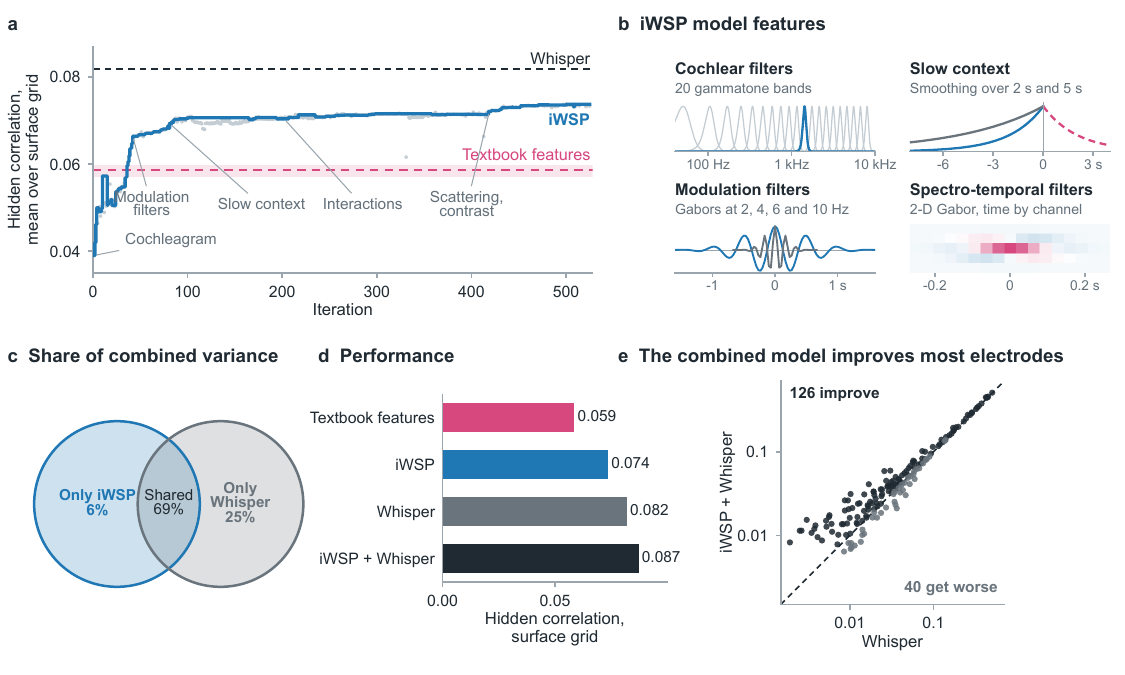}
  \caption{\textbf{Written audio features predict ECoG well and predict signal that Whisper fails to capture.} \textbf{a}, On the hidden set, using the grid electrodes, iWSP passes textbook features early in the search but stays below Whisper. Dots show each iteration, the line follows the model the search kept (its best on the visible set so far), and the band spans five textbook feature spaces that Claude Opus 5.5 wrote in one shot without feedback, with their mean dashed.
  \textbf{b}, The features underlying the iWSP model are combinations of known, interpretable kernels, e.g.  a filter bank, smoothing over the past 2\,s (blue), past 5\,s (gray) and next 2\,s (dashed), Gabors at 2, 4, 6 and 10\,Hz, and a 2-D Gabor. \textbf{c}, On the hidden set, these features capture variance that Whisper does not. Each region gives the share of the combined model's variance explained only by iWSP, by both, or only by Whisper, taken as the mean signed squared correlation over all 166 electrodes. \textbf{d}, Hidden-set correlation on the surface grid, where the textbook features were scored; combining iWSP and Whisper in one ridge regression improves on Whisper alone.  \textbf{e}, This improvement happens across many electrodes. Each point is one electrode's hidden-set correlation, on log scales; points above the dashed line improve when the iWSP features are added to Whisper.}
  \label{fig:ecog}
\end{widefig}
\end{figure}

\paragraph{iWSP on its own slightly underperforms Whisper.}
On the hidden set, using the grid electrodes, iWSP starts at 0.039 with a plain gammatone cochleagram and ends at 0.074, below the 0.082 of Whisper (\cref{fig:ecog}a).
The 3{,}852 features in the final model are combinations of known, interpretable kernels, and each gain comes from hand-implementing the following: compression and adaptation, spectro-temporal modulation filters\citep{chi2005multiresolution}, slow context over the previous seconds, and wavelet scattering\citep{anden2013deep} (\cref{fig:ecog}b). On the hidden set, over all 166 electrodes, iWSP reaches 0.065 against 0.070 for Whisper and leads on 77 electrodes. A 197-word description of these features is enough for fresh agents to rebuild the model with nearly the same hidden-set performance (\cref{fig:ecog_bottleneck}).
The search goes beyond the textbook features: the iWSP model passes them early in the search, while it adds the modulation filters, and stays well above them (\cref{fig:ecog}a, d). It also outperforms them on all five more patients described in \cref{sec:app_ecog}.

\paragraph{iWSP predicts signal that Whisper misses.}
On the hidden set, the iWSP features capture unique variance that Whisper does not: they alone explain 6\% of the combined model's variance, Whisper alone explains 25\% and the two share 69\% (\cref{fig:ecog}c).
As a result, combining both in one ridge regression significantly improves on Whisper alone, raising performance from 0.070 to 0.075 (this gain of 0.0052 has a 95\% interval of 0.0039 to 0.0064 from a bootstrap over electrodes).
The improvement happens across many electrodes (\cref{fig:ecog}e): 126 of 166 improve and 40 get worse. On five more patients, each evaluated on a hidden recording session, Whisper outperforms the iWSP model on every patient, but the iWSP features still explain variance that Whisper misses in four of the six (\cref{fig:ecog_multisubject}).

\paragraph{Which features are important.}
Because every feature has a formula, we can ask which ones the model relies on (\cref{fig:ecog_varpart}). On the visible set, the feature families overlap heavily, and 87\% of the variance explained in the brain responses cannot be assigned to any one of them. The cochleagram with slow context and a set of modulation filters predict the most unique variance, and one feature, spectral flatness, reaches 29\% of the full model alone.

\FloatBarrier

\section{Related works}
\label{sec:related}

\paragraph{Inherently interpretable models.}
A long line of work aims to build highly accurate models without sacrificing interpretability\citep{rudin2019stop,singh2021imodels}.
In text, this began with word-level and ngram-level models\citep{huth2016natural,lewis1998naive,bojanowski2016enriching,joulin2016bag}, which have continued to improve by using modern techniques to improve their construction\citep{liu2024infini,kim2024interpretable,singh2023augmenting}.
At the same time, concept-based models have become popular by routing predictions through a bottleneck of interpretable concepts\citep{koh2020concept,sun2024concept,benara2024crafting,team2026scaling,feng2026human,ludan2023interpretable,tan2024interpreting}.
However, the constrained form of these models has struggled to make accurate predictions for many hard problems, which iWSP aims to solve.

\paragraph{Agents that search for programs.}
Iterative coding agents offer a way around constrained model forms, because they can search a far larger space of readable models than traditional interpretable models can.
AlphaEvolve and related works\citep{romera2023mathematical,novikov2025alphaevolve,georgiev2025mathematical} iteratively write programs to solve particular problems in domains like math, and have been applied in domains like human and animal learning from data\citep{kasenberg2026ai}, or more broadly in automating the entire scientific pipeline\citep{lu2024ai,gottweis2025towards,swanson2025virtual,ifargan2024autonomous,tang2025ai,mitchener2025kosmos} (although benchmarks and critiques question their reliability\citep{chen2024scienceagentbench,luo2025more}).
iWSP follows a similar loop to these methods, but instead focuses on writing model weights and features.

\paragraph{Translating programs into weights.}
The systems above write programs rather than neural network weights. A separate line of work connects the two by compiling programs into networks. RASP is a small language whose primitives map onto attention and MLP operations\citep{weiss2021thinking}, and Tracr compiles RASP programs into transformer weights\citep{lindner2023tracr}. Learning Transformer Programs goes the other way, training transformers constrained to decompile into readable code, or decompiling them afterward\citep{friedman2023learning,thurnherr2024neural,huang2026discovering}. Unlike these methods, iWSP produces networks without an intermediate domain-specific language or compiler.

\paragraph{Networks that generate weights directly.}
Weights can also be written by another network instead of an agent. Hypernetworks are networks that write the weights of other networks\citep{ha2016hypernetworks,chauhan2024brief,han2026survey}, and HyperNEAT evolves a compact network whose outputs encode the weights of a much larger one\citep{stanley2009hypercube}.
Later work generates weights from collections of trained models, by sampling from learned representations of model zoos\citep{schurholt2022hyper}, by running a diffusion model on checkpoints\citep{peebles2022learning} or by predicting parameters for unseen architectures\citep{knyazev2021parameter}. Weight-space architectures respect the symmetries of these inputs\citep{navon2023equivariant}.
Like iWSP, these methods can set a target network's weights without running gradient descent on it, but they generally fail to provide explanations for their weight choices.

\paragraph{Post-hoc interpretation.}
All of the work above builds models, whereas the related field of post-hoc interpretability explains models that are already trained. Attribution methods estimate how much each input contributes to a prediction\citep{ribeiro2016should,lundberg2017unified,selvaraju2017grad,singh2019Hierarchical}. Mechanistic interpretability instead reads a model's internals\citep{elhage2021mathematical,geva2021transformer,olsson2022context,wang2022interpretability,merullo2023circuit,lieberum2023does,wu2024retrieval,gao2024scaling} and reverse-engineers small trained models\citep{nanda2023progress,quirke2023understanding}. LLMs and agents now automate much of this work by explaining in language what parts of a network do\citep{bills2023language,singh2023explainingmodules,hernandez2022natural,paulo2024automatically,shaham2024multimodal,liu2026neuronscope,marinllobet2026automated,zaigrajew2026line,khan2026can}.
In contrast, iWSP starts from explanations and uses these to implement and test weights.

\section{Discussion}

The main advance in this work is a method to build models that are both predictive and interpretable: instead of training a model and explaining it afterward, an agent writes it by hand and revises it based on its measured performance.
By applying iWSP across domains, we show that it can produce readable transformers for algorithmic tasks, and fairly accurate encoding models in predicting brain responses to different modalities.
More broadly, iWSP materializes explicit hypotheses as model weights that can be read, tested and revised, offering a path to align the explanations that science seeks with the predictive power of data-driven models.

In neuroscience specifically, the field has always been concerned with finding the features of stimuli that selectively drive activity in the brain.
Traditionally this problem has been approached by manually searching for these stimulus properties across multiple experiments\citep{kanwisher2006fusiform,kamps2016occipital} while data-driven methods\citep{jain2018incorporating,toneva2019interpreting,abnar2019,goldstein2022shared,antonello2023scaling} have explored a much larger model space, but sometimes failed to translate their results to human-understandable findings.
iWSP enables a more grounded, prior-based search to draw interpretable conclusions from data.

An important ancillary benefit of iWSP in neuroscience will be to help make datasets usable for future data-driven discovery. The signal we found at the start and end of each scan in the language dataset (\cref{sec:story}) echoes the usefulness of earlier work on inherently interpretable models.
For example, in a past study\citep{caruana2015intelligible}, a generalized additive model predicting pneumonia risk identified that asthma lowered predicted risk of death, because patients with asthma received more aggressive care, helping identify issues with confounding that can be used to correct a model.
As data-driven methods become increasingly powerful, interpretability will be more important in identifying and mitigating such confounds (and reward hacking more generally\citep{amodei2016concrete,skalse2022defining,bergen2026monitoring}).

\paragraph{Limitations.}
Despite these strengths, iWSP has important limitations in both the models it discovers and the evidence they provide. First, current agents do not invent genuinely new computational principles. Instead, they explore and combine known motifs drawn from their prior knowledge, such as textbook algorithms on synthetic tasks (\cref{sec:wsp}) or established cortical operations like Gabor filters, divisive normalization, and modulation filters on brain data (\cref{sec:vision,sec:ecog}). Systematically combining known motifs is already productive, but proposing creatively distinct, unfamiliar hypotheses remains beyond current models. Relatedly, the discovered models rely on shallow motifs rather than deep, multi-layer hierarchies: our brain encoding searches assemble parallel banks of one- or two-stage computations into a linear readout, and our NLP transformers stack only a few rule-based stages. Programming deep hierarchical representations without backpropagation remains an open problem, which likely explains why iWSP matches pretrained networks in early visual cortex while trailing deeper models in ventral visual cortex, harder language tasks, and speech ECoG. At the method level, iWSP relies on an agent's written explanations remaining faithful to its weight edits; while our audits and text-bottleneck rebuilds (\cref{sec:wsp,sec:vision,sec:story,sec:ecog}) support this for current frontier models, stronger agents will require continued vetting to ensure they do not embed opaque heuristics. Running search on neural datasets is also slow and costly, requiring hundreds of candidate evaluations and a large token budget, although these costs continue to fall as LLMs become faster and cheaper. Finally, our neuroscience experiments are observational: they include relatively few subjects, test generalization within similar populations, and lack causal interventions. In visual cortex, for instance, area attributions and driving images reflect the specific hypotheses written into one \visModelLines{}-line model on the data that guided its search, showing which explicit computations predict responses rather than identifying unique causal mechanisms in the brain.

\paragraph{Future work.}
The work here opens the door to a great deal of future research.
One idea is to use iWSP to explain black-box models by approximating them with an interpretable network: this could help build a mechanistic understanding of when they work/fail, improving over modern toy models used in mechanistic interpretability\citep{elhage2022toy,nanda2023progress,chughtai2023toy,quirke2023understanding,lindner2023tracr}.
In data-scarce domains (often the case in neuroscience and medicine), another path
is to leverage the explainability offered by iWSP to allow domain experts to help improve model adaptation\citep{amershi2014power,holzinger2016interactive,ross2017right,teso2019explanatory,feng2026human,schramowski2020making}.
Finally, iWSP could also serve as an automated audit of widely used datasets, and could be applied to a much broader range of domains and tasks, including other neuroscience modalities and beyond.
The capability of frontier models to do these weight-level tasks is exciting for domains like science, and is highly verifiable, making it possible to rapidly improve at this capability.
We hope iWSP will help move toward a future in which the models we rely on for science are ones that we can interpret and trust.

\renewcommand{\bibfont}{\footnotesize}
\setlength{\bibsep}{3pt plus 1pt}
\bibliographystyle{unsrtnat}
\FloatBarrier
{\small
\bibliography{refs}
}
\FloatBarrier

\appendix
\setlength{\intextsep}{6pt plus 2pt minus 2pt}
\setlength{\floatsep}{8pt plus 2pt minus 2pt}
\setlength{\textfloatsep}{10pt plus 2pt minus 4pt}
\captionsetup{skip=5pt}
\section*{\centering Appendix}
\renewcommand{\thetable}{A\arabic{table}}
\renewcommand{\thefigure}{A\arabic{figure}}
\setcounter{table}{0}
\setcounter{figure}{0}
\renewcommand{\thesubsection}{A.\arabic{subsection}}

\makeatletter
\let\iwsp@addcontentsline\addcontentsline
\renewcommand{\addcontentsline}[3]{\iwsp@addcontentsline{#1}{#2}{#3}%
  \def\iwsp@ext{#1}\def\iwsp@toc{toc}\ifx\iwsp@ext\iwsp@toc\iwsp@addcontentsline{apc}{#2}{#3}\fi}
\noindent\textbf{Contents}\par
\@starttoc{apc}
\makeatother
\bigskip

\subsection{Transformers for algorithmic and NLP tasks}
\label{sec:app_wsp}

\begin{table}[!ht]
  \centering
  \caption{\textbf{The 17 tasks.} Models read the prompt on the left and generate the answer on the right one token at a time. For NLP tasks we show the input format, and we shorten the padding in date normalization.}
  \label{tab:tasks}
  \footnotesize
  \input{tables/wsp_task_list}
\end{table}

\begin{table}[!ht]
  \centering
  \caption{\textbf{Audit results and hidden accuracy for each LLM across the 17 tasks.} We classify each iteration by rerunning its weight-writing code on altered task data, and count an iteration as not checked if its code could not be rerun or was not deterministic. Hidden accuracy is measured on the first model to reach each run's best visible accuracy. The size row only includes checked models; including all iterations gives medians of 166, 35{,}850 and 100{,}000. We added two sentences to the Sonnet and Haiku prompts on how to learn the task format, since their first sessions stopped early.}
  \label{tab:wsp_models}
  \footnotesize
  \input{tables/wsp_model_comparison}
\end{table}

\begin{figure}[!ht]
  \centering
  \includegraphics[width=.55\linewidth]{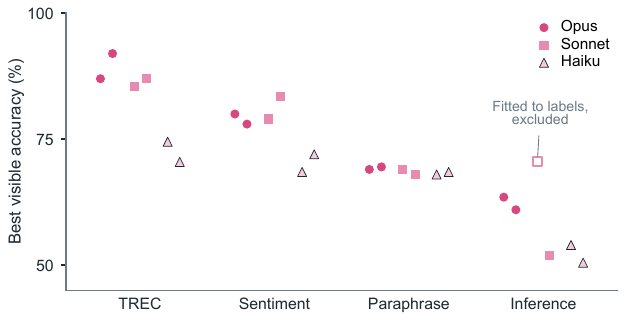}
  \caption{\textbf{NLP tasks across Claude models.} We show the best visible accuracy of each run, counting only models whose weights do not depend on task data. The open square marks the Sonnet run that fit its models to labels, which we exclude.}
  \label{fig:wsp_models}
\end{figure}

\FloatBarrier

\begin{figure}[!ht]
  \centering
  \includegraphics[width=\linewidth]{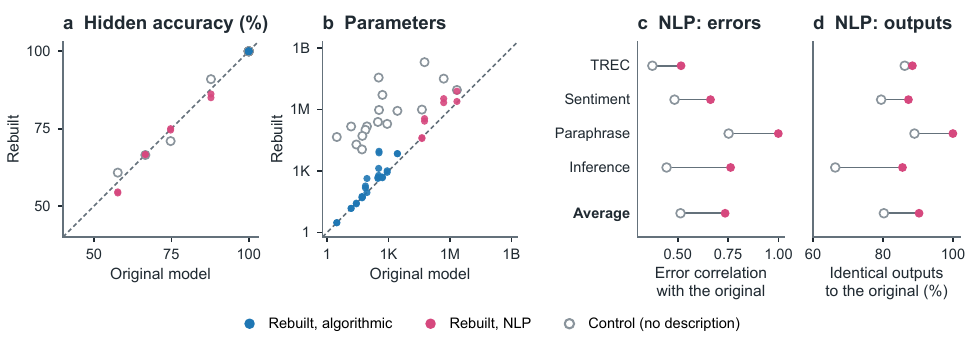}
  \caption{\textbf{Rebuilding models from 200-word descriptions.} For each of the 17 tasks, the best Opus 5.5 model is described in at most 200 words and rebuilt twice from that description (filled; blue, algorithmic; pink, NLP). A control session rebuilds it with no description (open). Without a description to follow, the control is not constrained to a compact or interpretable model, so it often writes far larger ones, such as hand-written rule lists with 207 million parameters for TREC that can score better than the original. Dashed lines mark equality with the original.
  \textbf{a}, Hidden accuracy. All 13 algorithmic originals, rebuilds and controls reach 100\%.
  \textbf{b}, Parameters, on log scales. Rebuilt algorithmic models have 1.74 times the original's parameters on a geometric mean, against 2{,}043 times for the control.
  \textbf{c}, \textbf{d}, How closely each model reproduces the original's per-example behavior on the NLP hidden sets: the correlation between its errors and the original's, and the share of outputs identical to the original's, averaged over the two rebuilds. Algorithmic tasks are not shown because their original, rebuilt and control models are all exact, so they make no errors to correlate and their outputs are always identical. Averaged over the NLP tasks, the rebuilds' error correlation exceeds the control's by 0.22 (95\% bootstrap interval over examples, 0.18 to 0.26). TREC has the weakest evidence because its hidden set has only 500 examples.}
  \label{fig:wsp_bottleneck}
\end{figure}

{\scriptsize
\setlength{\LTleft}{-.5in}\setlength{\LTright}{0pt plus 1fill}\setlength{\LTcapwidth}{\dimexpr\textwidth+1in\relax}
\begin{longtable}{@{}p{.9in}p{\dimexpr\textwidth+1in-.9in-2\tabcolsep\relax}@{}}
\caption{\textbf{Six of the 17 descriptions given to the rebuilding sessions, chosen at random.} Claude Opus 5.5 wrote each one from the model's code, and we reproduce them word for word, with word counts in parentheses. The GCD model here is the best Opus 5.5 model, which is more compact than the GCD model in \cref{fig:params}e. All 17 are \href{https://github.com/csinva/weight-space-programming-autoresearch}{on GitHub}.}
\label{tab:wsp_descriptions} \\
\toprule
Task & Description \\
\midrule
\endfirsthead
\toprule
Task & Description \\
\midrule
\endhead
\bottomrule
\endlastfoot
\input{tables/wsp_bottleneck_descriptions}
\end{longtable}
}

\FloatBarrier

\subsection{Computations of the visual cortex model}
\label{sec:app_visual_model}

\paragraph{The iWSP model cascades a V1-like filter bank into five explicit mid-level modules.}
The iWSP model is a Python program of \visModelLines{} lines with no learned weights that mirrors the hierarchical organization of biological vision. At a high level, the model first passes the image through a shared V1-like frontend to extract normalized local edge and contrast maps, branches into five parallel mid-level modules that capture higher-order structure, and finally pools these maps over a foveated spatial grid. In the frontend, pixel intensities undergo power-law compression and are split into one luminance and two color-opponent channels (red--green and blue--yellow). Each channel is convolved at three spatial scales with center--surround filters and oriented Gabor filters spanning eight orientations. The responses are then passed through standard cortical nonlinearities: quadrature Gabor pairs are combined into phase-invariant complex-cell energy\citep{adelson1985spatiotemporal} (the square root of summed squared responses), center--surround responses are half-wave rectified into ON and OFF channels, and all channels undergo local divisive normalization followed by logarithmic compression to encode contrast relative to the surrounding image context.

From this normalized V1 representation, the model computes five kinds of mid-level features:
(i) \emph{texture boundaries}, which pass the first-stage energy maps through a second bank of Gabor filters and energy pooling (a classical filter--rectify--filter cascade) to detect where patterns of oriented contrast change;
(ii) \emph{contour curvature}, which takes spatial derivatives of the local orientation field to measure how rapidly edge orientation bends across space and relative to its surround;
(iii) \emph{edge phase}, which compares even and odd Gabor responses rectified by sign to distinguish thin bright ridges, dark creases, and step edges between surfaces;
(iv) \emph{feature co-occurrences}, which multiply pairs of centered, normalized channels and average them spatially to measure which orientations, colors, curvatures, and phases appear together within and across spatial scales, following classical parametric texture models\citep{portilla2000parametric}; and
(v) \emph{surface geometry}, which acts directly on smoothed luminance and color derivatives to classify local brightness surfaces into five 3D shading primitives (pits, valleys, saddles, ridges, and bumps) alongside corner strength and isophote curvature.
\Cref{fig:fmri}d shows the weights and operations of five of the model's computations.
Finally, every feature map is summarized by its mean and spatial standard deviation over a foveated grid of image windows (finer near the center of gaze) and normalized per block, yielding 9{,}996 explicit features. Aside from the analytic tuning curves used to categorize the five surface shapes, nearly every nonlinearity in the program reduces to one of five canonical operations from visual neuroscience: power/log compression, half-wave rectification, quadrature energy, divisive normalization, and pairwise channel multiplication.

\begin{figure}[ht]
  \centering
  \includegraphics[width=\linewidth]{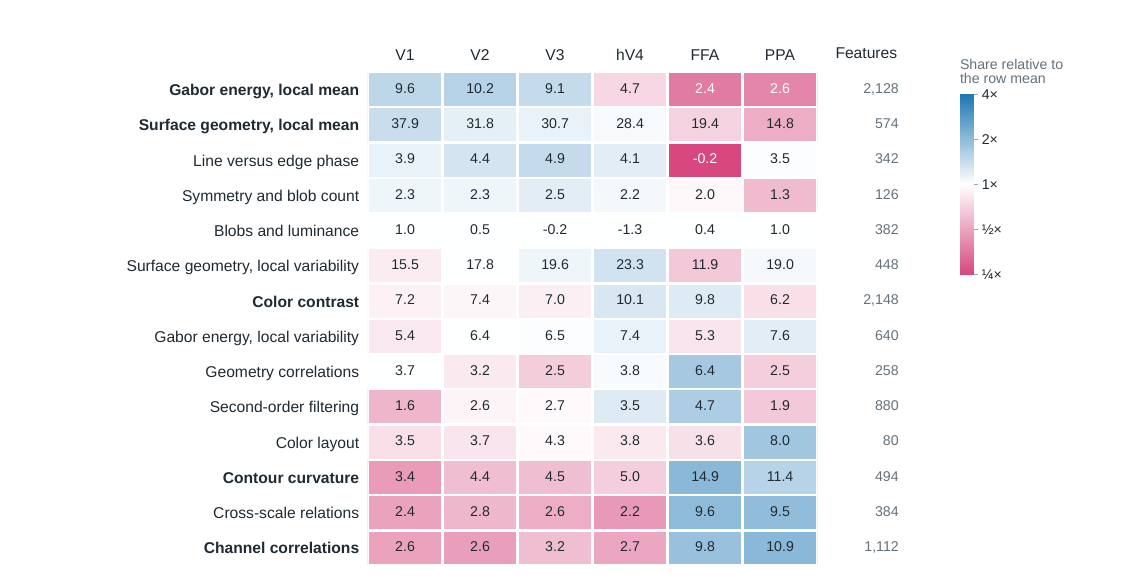}
  \caption{\textbf{Share of each area's prediction carried by each of the model's 14 computations.} Cells give the share in percent, averaged over five LAION-fMRI subjects; the shares in a column sum to 100\%. Color shows a computation's share in an area relative to its mean share over the six areas, and the one row with a mean share below 1\% is left uncolored. Rows are ordered from computations that favor early areas to those that favor FFA and PPA. Bold rows appear in \cref{fig:fmri}d. The last column gives the number of output features of each computation.}
  \label{fig:fmri_attribution}
\end{figure}

\paragraph{Attributing each area's prediction to computations.}
Each of the 9{,}996 output features is written by one stage of the program, and we group the features into 14 computations by that stage. Where a stage summarizes a map both by its average within the cells of a spatial grid and by its variability within those cells, we treat the two summaries as separate computations. \Cref{fig:fmri_attribution} lists the 14 computations with their feature counts. To attribute an area's prediction, we fit the ridge regression once with all features, as in the search. For one voxel, the prediction on the shared images is a sum over computations, $\hat{y} = \sum_c \hat{y}_c$, where $\hat{y}_c$ uses only the features and ridge weights of computation $c$. The correlation between the voxel's prediction and the measured response $y$ then splits exactly into one term per computation:
\[
  \mathrm{corr}(\hat{y}, y) = \sum_c \frac{\mathrm{cov}(\hat{y}_c, y)}{\sigma_{\hat{y}}\,\sigma_{y}}.
\]
We average each term over the voxels of an area, divide by the area's mean correlation so that the shares sum to 100\%, and then average over the five LAION-fMRI subjects. A share is negative when a computation's part of the prediction is anticorrelated with the response. Correlated computations divide the credit between them, and the model scales its feature blocks differently, so a share describes what the fitted ridge regression uses, not what it needs. We therefore also measure each computation's unique variance: we refit ridge without the computation's features and report the drop in squared correlation as a percentage of the full model's squared correlation (\cref{tab:fmri_unique}). The maps in \cref{fig:fmri}d use the finest image scale, and the matrix of channel correlations uses half that resolution, as in the model. The five surface-shape tiles in \cref{fig:fmri}d illustrate the brightness profiles that the code separates and are not filter weights. The matrix of channel correlations has $24 \times 24$ entries, with positive values in blue and negative values in orange.

\begin{table}[ht]
  \centering
  \caption{\textbf{Unique variance of each computation of the visual cortex model.} Values give the drop in squared correlation when ridge is refit without one computation, as a percentage of the full model's squared correlation in each area (mean $\pm$ standard error over five LAION-fMRI subjects). A negative value means that the ridge regression predicts the area better without the computation. The last row gives the variance that is not unique to any one computation.}
  \label{tab:fmri_unique}
  \footnotesize
  \input{tables/fmri_unique_variance}
\end{table}

\paragraph{Search constraints and held-out evaluation.}
In every candidate, each filter and weight is set analytically or generated from a fixed random seed. Candidates are limited to fewer than 10{,}000 output features so that a model cannot succeed merely by handing the ridge regression an overcomplete random basis. On NSD, we evaluate the \visNsdSubjects{} subjects who completed all scanning sessions, fitting the linear readout on each subject's unique images and testing on the shared images after excluding any stimulus that also appears in LAION-fMRI, which leaves \visNsdImages{} test images. The seven interpretable baselines are Gabor filters, Gabor filters with divisive normalization\citep{heeger1992normalization}, GIST\citep{oliva2001modeling}, HMAX\citep{riesenhuber1999hierarchical}, wavelet scattering\citep{bruna2013invariant}, Portilla--Simoncelli texture statistics\citep{portilla2000parametric}, and a contour curvature model. For AlexNet, we use for each stream the layer that scores best on the \visSearchSubjects{} search subjects and visible images of LAION-fMRI, averaged over subjects and fixed before any held-out evaluation: \visAlexEarlyLayer{} for early and \visAlexVentralLayer{} for ventral visual cortex. The same two layers are used for held-out LAION-fMRI and for NSD. For the combined baseline, we concatenate the features of the seven baselines (\visCombinedFeatures{} features in total) and fit the same ridge regression, with one penalty per voxel. Before concatenation, we center each baseline's features and divide them by their root mean squared norm over the training images, so that the seven baselines enter the regression with equal weight; without this rescaling, the combined baseline scores lower in ventral visual cortex on both datasets. Across individual subjects, the iWSP model exceeds the strongest single baseline in every NSD subject by \visNsdEarlyGapMin{} to \visNsdEarlyGapMax{} in early visual cortex and by \visNsdVentralGapMin{} to \visNsdVentralGapMax{} in ventral visual cortex. On the \visFinalSubjects{} held-out LAION-fMRI subjects, the best model of the \visSearchSubjects{}-subject ablation search reaches \visLaionPooled{} pooled correlation (\visLaionEarly{} versus \visLaionEarlyBest{} for \visLaionEarlyBestName{} in early visual cortex, higher by \visLaionEarlyGapMin{} to \visLaionEarlyGapMax{} per subject; \visLaionVentral{} versus \visLaionVentralBest{} for \visLaionVentralBestName{} in ventral visual cortex, higher by \visLaionVentralGapMin{} to \visLaionVentralGapMax{} per subject). On the \visSearchSubjects{} search subjects of that run, the model scores identically on held-out and visible shared images (\visSearchSubjectsHidden{} versus \visSearchSubjectsVisible{} pooled correlation), showing that the lower absolute correlation on the \visFinalSubjects{} held-out subjects reflects subject differences that affect baselines as well (Gabor filters with divisive normalization fall from \visGaborNormSearch{} on the search subjects to \visGaborNormHeldOut{} on the held-out subjects).

\begin{figure}[ht]
  \centering
  \includegraphics[width=\linewidth]{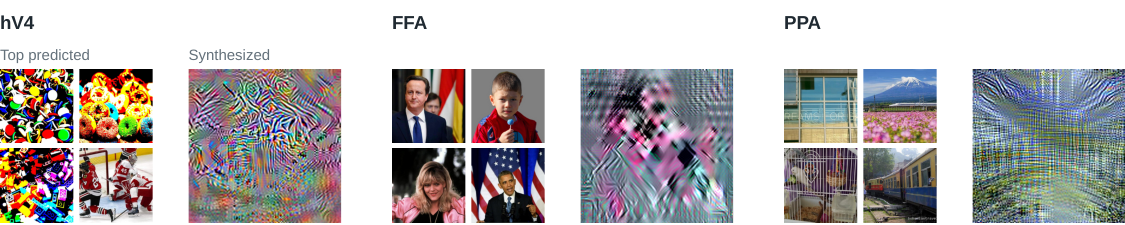}
  \caption{\textbf{Images that drive the model's predictions for hV4, FFA and PPA.} For each area, the four of the 1{,}492 shared LAION-fMRI images with the highest predicted response are shown next to an image synthesized from noise to maximize the predicted response. Synthesized images have a fixed contrast and reach predictions beyond those for any natural image, so they are extrapolations.}
  \label{fig:fmri_drivers}
\end{figure}

\paragraph{Area readouts and synthesized images.}
To see which images drive the model's prediction for each area (\cref{fig:fmri_drivers}), we fit the ridge readout for every voxel of an area as in the search, average the voxel weight vectors within the area, and then average over the five LAION-fMRI subjects. The resulting area readout predicts the area-mean response, and the shared images are held out of every readout fit. It predicts the measured area-mean response to the shared images with a correlation of 0.82 in hV4, 0.53 in FFA, and 0.70 in PPA. The images with the highest predicted response are dense, multicolored objects for hV4, faces and people for FFA, and buildings and other rectilinear scenes for PPA; the 20 highest-ranked images of each area evoke measured responses 1.3 to 1.7 standard deviations above the area's mean. Because the model is differentiable, we can also synthesize an image that maximizes each area readout. To synthesize an image, we start from noise and run gradient ascent on the area readout, with the image parametrized in the Fourier domain and shifted by up to 6 pixels at random on each step. We fix the image mean at 0.5 and the pixel standard deviation at 0.2, because without this constraint the optimum is a blank image. Two blocks of global image statistics (82 features) have no usable gradient and are held constant. The synthesized images reach predicted responses 6.5 (hV4), 9.6 (FFA), and 10.3 (PPA) standard deviations above the mean prediction for the shared images, where the highest natural image reaches about 2.5, so they show the direction the readout favors rather than a response the model has been tested on. The synthesized images agree with the natural ones: fine multicolored texture for hV4, a central skin-toned blob for FFA, and a lattice of horizontal and vertical lines for PPA. These images show which image statistics the model associates with each area; they do not show that the model represents faces or scenes as categories.

\begin{figure}[!ht]
  \centering
  \includegraphics[width=\linewidth]{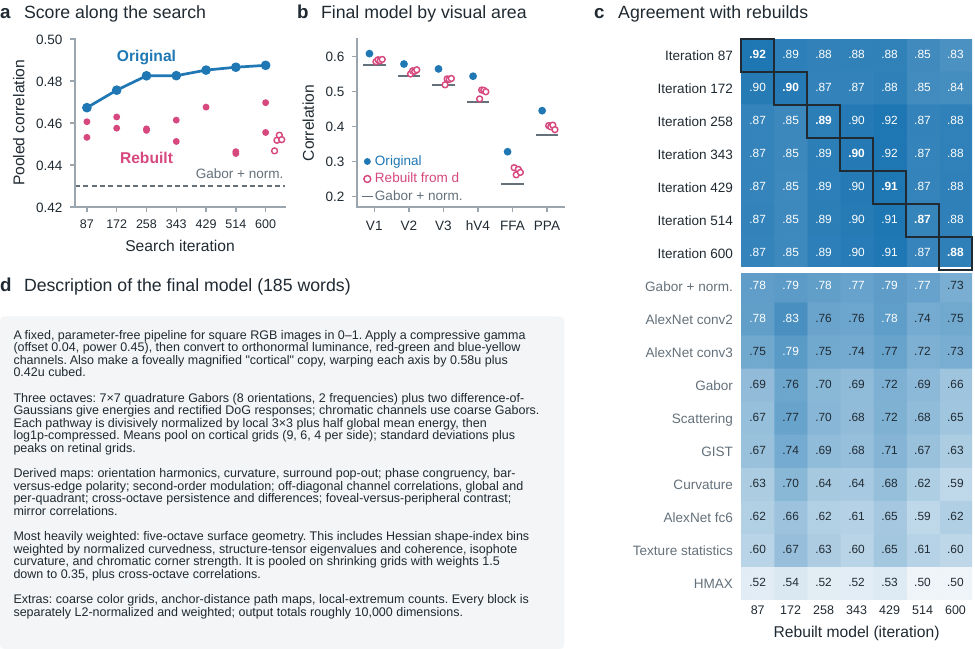}
  \caption{\textbf{The visual cortex model after a text bottleneck.} A model is described in at most 200 words and rebuilt from that description and the task, as in \cref{fig:loop}b. Each score averages voxelwise Pearson correlations within a region and then across five LAION-fMRI subjects. Readouts are fit on each subject's unique images and evaluated on the shared visible images; these are not held-out-subject evaluations.
  \textbf{a}, Pooled encoding correlation of the search's best model at eight iterations (blue) and of two rebuilds of each model (pink dots). The dashed line marks normalized Gabor features. Open circles show the four rebuilds of the final model from the description in \textbf{d}. One rebuild at iteration 429 fails to run and is not shown. The model at iteration 1 and both of its rebuilds score 0.148, below the range of the axis.
  \textbf{b}, Correlations of the final model and of its four rebuilds from the description in \textbf{d} in six visual areas. Gray lines mark normalized Gabor features.
  \textbf{c}, Correlation between the predicted responses of a model (rows) and those of the rebuilds of each search model (columns) to the 1{,}492 shared images, computed per voxel and averaged over all voxels, over the rebuilds of a model, and over the five subjects. Rows list the seven search models, then seven classical baselines and three AlexNet layers ordered by mean correlation with the rebuilds. The outlined diagonal pairs each search model with its own rebuilds.
  \textbf{d}, The 185-word description of the final model, which Claude Opus 5.5 wrote from the model's code, reproduced word for word.}
  \label{fig:fmri_bottleneck}
\end{figure}

\paragraph{Rebuilding the visual cortex model from text.}
\label{sec:app_vision_bottleneck}
To test whether a short description preserves the iWSP model's computations, Claude Opus 5.5 reads the model's code and summarizes it in 185 words (\cref{fig:fmri_bottleneck}d). We give only that description and the task to four independent rebuilding sessions. Each rebuilt feature extractor receives a newly fitted ridge readout under the original five-subject LAION-fMRI protocol. Mean pooled correlation is 0.451 for the rebuilds and 0.488 for the original, and in each of six visual areas every rebuild scores below the original and above normalized Gabor features (\cref{fig:fmri_bottleneck}a, b). The rebuilds recover 37\% of the original's correlation gain over normalized Gabor features (correlation 0.430), indicating partial reconstruction of the improvement beyond this baseline.

Similar accuracy does not show that the rebuilds compute the same thing, so we also compare what the models predict. For each voxel, we correlate two models' predicted responses to the 1{,}492 shared images. The rebuilds' predictions correlate with the original's at 0.86 to 0.88, compared with 0.81 for pretrained AlexNet conv2, a model of similar accuracy (correlation 0.461), and 0.80 for normalized Gabor features. Because every model shares a large Gabor-like component, we repeat the comparison after regressing each prediction on the normalized Gabor prediction, voxel by voxel. On this remainder, the rebuilds correlate with the original at 0.64 to 0.69, against 0.50 for AlexNet conv2, and every rebuild exceeds AlexNet conv2 in each of the five subjects. Over each subject's 10\% of voxels best predicted by the original, the original correlates with the rebuilds at 0.94 (0.74 to 0.76 after the regression) and with AlexNet conv2 at 0.91 (0.57 after the regression). The description therefore carries computations specific to the original, but only in part: the rebuilds agree more with one another (0.77 to 0.83) than with the original, so they share a reading of the text that differs from the code. Two further comparisons do not favor the rebuilds. Across voxels, the gain of each model over normalized Gabor features correlates with the original's gain at 0.61 to 0.73 for the rebuilds and at 0.73 for AlexNet conv2. The pattern of per-voxel accuracy correlates at 0.96 or above for every pair of models, including both references, so it does not distinguish them. Linear centered kernel alignment between original and rebuilt features ranges from 0.71 to 0.76 on 1{,}000 test images. The rebuilds are also smaller than the original, with 292 to 358 lines of code against 742 and about a third of its numeric constants (\cref{tab:fmri_bottleneck_size}).

\begin{table}[!ht]
  \centering
  \caption{\textbf{Size of the original and rebuilt visual cortex models.} Lines of code count non-blank, non-comment lines (the original file has \visModelLines{} lines). Numeric constants count the numbers written in the code other than 0 and 1. Output features are the columns passed to the ridge readout. Pooled correlation is as in \cref{fig:fmri_bottleneck}a.}
  \label{tab:fmri_bottleneck_size}
  \footnotesize
  \input{tables/fmri_bottleneck_complexity}
\end{table}

\paragraph{The text bottleneck along the search.}
The comparison above uses only the final model, so we repeat the text bottleneck on the search's best model at eight evenly spaced iterations: 1, 87, 172, 258, 343, 429, 514, and 600. As before, Claude Opus 5.5 describes each model in at most 200 words (188 to 200 used), and two independent sessions rebuild each model from its description and the task alone. We score the originals and the rebuilds under the five-subject LAION-fMRI protocol of the search. One of the 16 rebuilds, at iteration 429, fails when run on the GPU and is left out.

The first model is a random projection of the pixels (pooled correlation of 0.148), and both of its rebuilds reproduce its score exactly. Across the eight models, the rebuilds keep 96\% of the original's performance on average. For the seven later models they keep between 92\% (iteration 514) and 98\% (iteration 87), and every rebuild outperforms normalized Gabor features (correlation 0.430; \cref{fig:fmri_bottleneck}a). No rebuild reaches its original: the originals improve from 0.467 at iteration 87 to 0.488 at iteration 600, while the rebuilds stay between 0.445 and 0.470. Over the same iterations the program grows from 373 to \visModelLines{} lines. A description of 200 words thus conveys most of each model, and the small improvements of the later iterations are the part it conveys least. The two rebuilds of the final model, which come from a new description, reach 0.456 and 0.470, in line with the four rebuilds above (0.447 to 0.454; open circles in \cref{fig:fmri_bottleneck}a).

We also ask whether the rebuilds of a model predict like their own original more than like the other checkpoints and standard baselines (\cref{fig:fmri_bottleneck}c). Comparing the seven sets of rebuilds against a 17-row reference set consisting of the seven later originals, the seven classical baselines, and AlexNet conv2, conv3, and fc6 requires no additional rebuilding runs. Each original correlates with its own rebuilds at 0.895 on average, versus 0.753 across all 112 off-diagonal pairs ($p = 4.7 \times 10^{-7}$, exact permutation test over all 98{,}017{,}920 ways of pairing seven of the 17 models with the seven sets of rebuilds). The ten baselines correlate with the rebuilds at 0.679 on average, from 0.520 for HMAX to 0.773 for normalized Gabor features and AlexNet conv2, and for every rebuilt checkpoint all seven search originals agree more closely with its predictions than any baseline does. Within the seven search checkpoints alone, off-diagonal agreement is 0.877 ($p = 0.009$ over all 5{,}040 pairings among the seven originals) because checkpoints from one search share most of their computations; from iteration 258 onward, the originals agree almost equally with each set of rebuilds, and an original agrees most with its own rebuilds for 3 of the 7 models.

\FloatBarrier
\clearpage
\subsection{Story-listening fMRI}
\label{sec:app_story}

\begin{figure}[!ht]
  \centering
  \includegraphics[width=\linewidth]{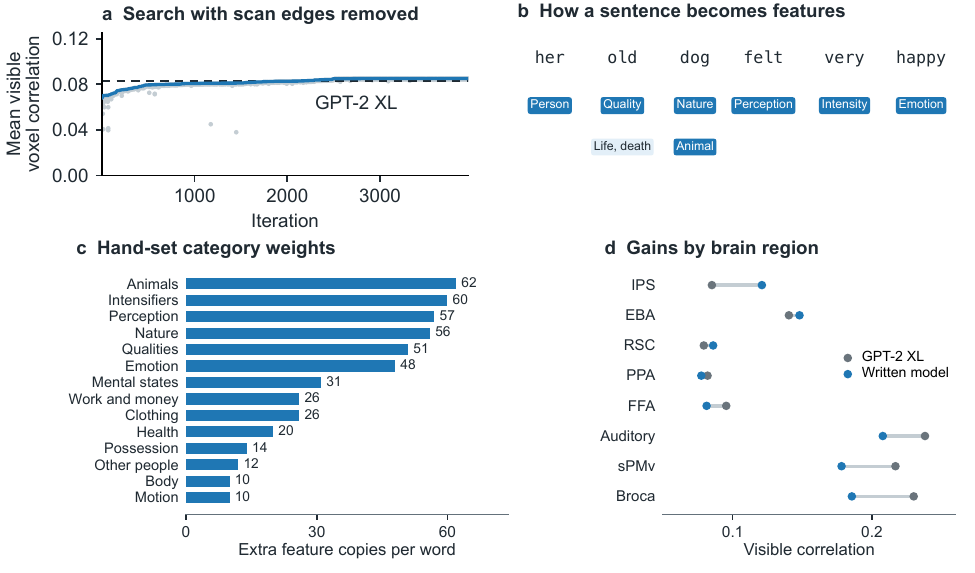}
  \caption{\textbf{With the start and end of each scan removed, a readable word model slightly passes GPT-2 XL.} \textbf{a}, We remove 30 TRs from each end of every scan and run a separate Claude Opus 4.7 search, which climbs gradually over 3{,}949 iterations with no jump and reaches 0.0851, against 0.0826 for GPT-2 XL. Gray points are iterations and the blue line is the best visible performance. \textbf{b}, The best model is a hand-written lexicon, shown here on a new sentence. It looks each word up in 43 semantic categories, then uses attention to pool these tags over each 10-word window at four time-scales. \textbf{c}, The agent hand-tuned the weight of each category, with the largest weights on categories such as animals, intensifiers and perception (blue in \textbf{b}). It also added 10 pairs and 23 triples of categories. \textbf{d}, The lexicon does not win everywhere: it beats GPT-2 XL in IPS, EBA and RSC but falls behind in auditory cortex and in the speech regions sPMv and Broca's area.}
  \label{fig:trimmed}
\end{figure}

\begin{figure}[!ht]
  \centering
  \includegraphics[width=\linewidth]{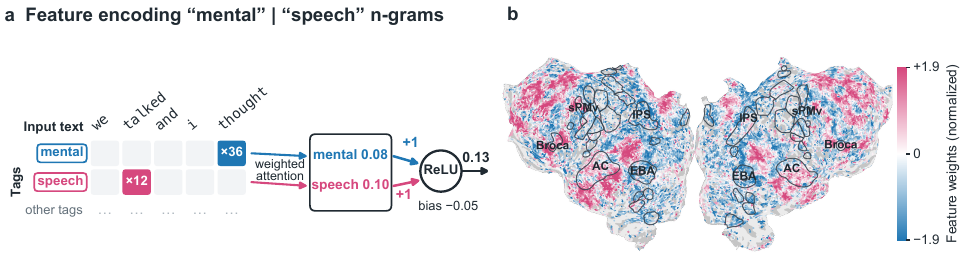}
  \caption{\textbf{One feature of a written language fMRI model and its ridge regression weights.} The feature comes from an earlier model of the iWSP search with the start and end of each scan removed (\cref{fig:trimmed}; visible performance 0.081) and encodes words related to mental processing or speech.
  \textbf{a}, The feature on the input text ``we talked and i thought''. The model tags each word using word lists written by iWSP, and each tag becomes a 1 in its own embedding dimension; rarer content words are upweighted, here $\times$36 for ``thought'' and $\times$12 for ``talked''. Weighted attention averages each tag over the 10-word window, with earlier words weighted more, giving shares of 0.08 for mental and 0.10 for speech words. A single ReLU neuron adds the two shares with weights of +1 and a bias of $-0.05$, so it outputs zero unless their sum passes 0.05; it outputs 0.13 here, and 0 for speech alone in ``she said it was over'' (share 0.049).
  \textbf{b}, Ridge regression weights of this feature for subject UTS03, summed over the four delays and normalized by each voxel's typical weight across all features, for voxels with visible correlation above 0.1. Each feature of the model enters ridge regression as its own column, so every feature can be mapped onto the cortex in the same way. The weight is positive in 61\% of the 10\% best-predicted voxels. Regions: auditory cortex (AC), Broca's area (Broca), superior premotor ventral (sPMv), extrastriate body area (EBA) and intraparietal sulcus (IPS).}
  \label{fig:story_feature}
\end{figure}

\begin{figure}[!ht]
  \centering
  \includegraphics[width=.8\linewidth]{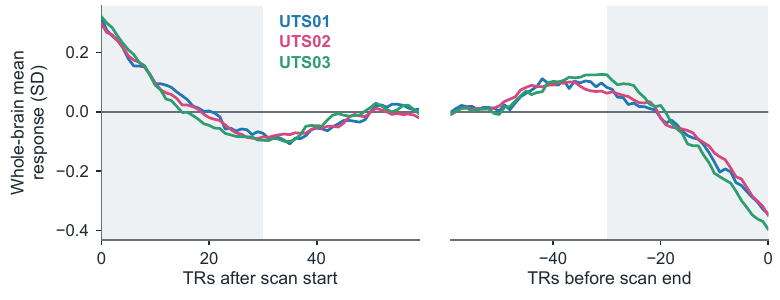}
  \caption{\textbf{The scan-timing signal appears in all three subjects.} We plot the mean response across all voxels, averaged over each subject's 84 story scans and aligned to the start and end of each scan, with the 30 TRs nearest each end shaded. In every subject, the first five TRs are above zero in at least 83 scans and the last five are below zero in at least 83 scans. The average curves correlate at 0.98 or higher between subjects.}
  \label{fig:scan_edges_subjects}
\end{figure}

\begin{table}[!ht]
  \centering
  \caption{\textbf{The performance jump due to the scan-timing features generalizes to the hidden set.} We evaluate GPT-2 XL and several models from the search on all TRs: the best model so far at roughly evenly spaced iterations, the last model before the jump, the first after it and the final model. Ridge regression is always fit on the search's 8 stories. The visible column gives the performance logged during the search, and the hidden column gives the mean voxel correlation on 10 other stories drawn at random. The last column repeats the hidden measurement without the 30 TRs at each end of every scan, with no refitting. Hidden performance is lower than visible performance for every model, including GPT-2 XL, because visible responses are averaged over repeated presentations of a story while each hidden story was presented once, not simply because of poor generalization.}
  \label{tab:story_hidden}
  \footnotesize
  \input{tables/story_hidden_set}
\end{table}

\begin{figure}[!ht]
  \centering
  \includegraphics[width=.85\linewidth]{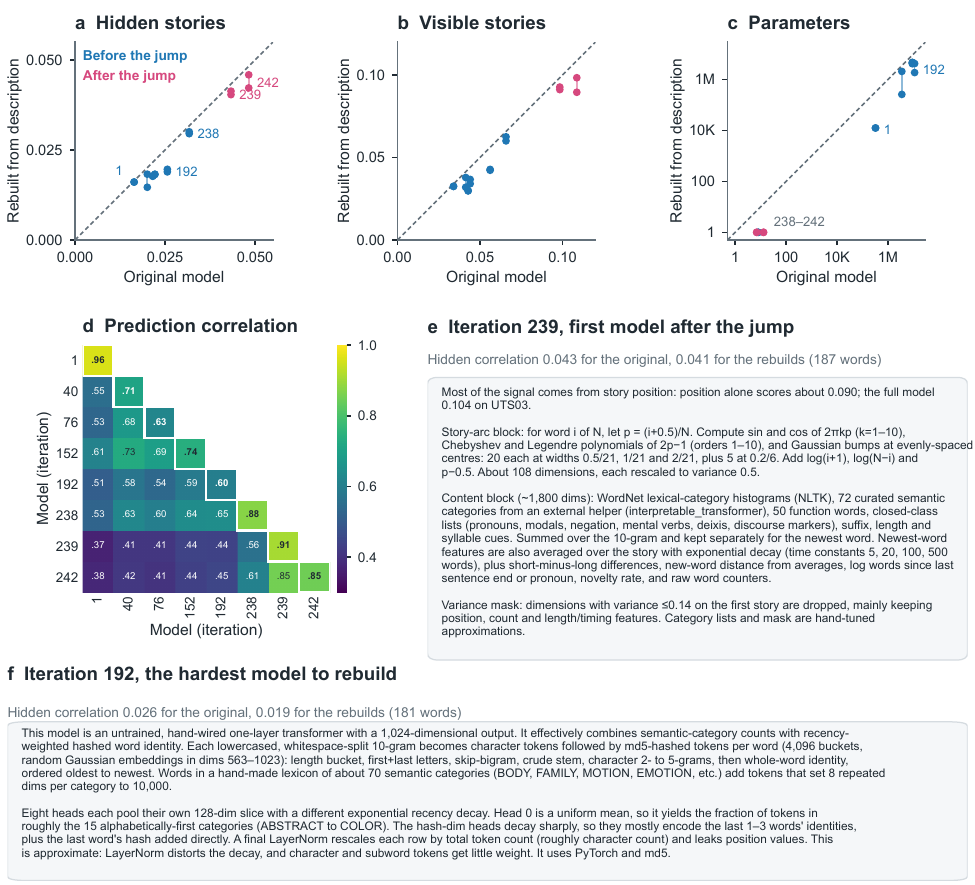}
  \caption{\textbf{A 200-word description is enough to rebuild the language fMRI models.} We apply the text bottleneck of \cref{fig:loop}b to the eight models of \cref{tab:story_hidden} from the iWSP search on all TRs: the first model (iteration 1), the best model so far at iterations 40, 76, 152 and 192, the last model before the jump (238), the first model after it (239) and the final model (242). A fresh Claude Code session with Claude Opus 5.5 reads each model's code and describes it in at most 200 words. Two further fresh sessions then rebuild each model from its description alone, under the same rules as the original search (no training and nothing fit to the fMRI responses), with at most 6 visible-set experiments each (to help ensure the network successfully compiles and adheres to the description) and no access to the hidden stories.
  \textbf{a}, \textbf{b}, Mean voxel correlation of each original and its two rebuilds on the hidden and visible stories, with ridge regression fit on the search's 8 fitting stories as in \cref{tab:story_hidden}; the dashed line marks equal performance. Rebuilds keep 87\% of the original's hidden performance on average, from 75\% for iteration 192 to 98\% for iteration 1, and both rebuilds of iteration 239 recreate its story-position features. Models from the middle of the search lose the most, since they rely on long hand-made word lists and hashed word identities that a description can name but not list.
  \textbf{c}, Parameters of each original and its two rebuilds, counted as every number a model stores (tensor and array elements and numeric constants; word lists are not counted), on log scales; counts of 0 are drawn at 1. Rebuilds of iterations 1 to 192 have 2 to 26 times fewer parameters than their originals, and the models from iteration 238 on, built from word lists, store at most 13 numbers.
  \textbf{d}, Correlation between the predicted hidden-story responses of each original and of the rebuilds of each model (averaged over the 2 rebuilds), computed per voxel and averaged over the original's 10\% best-predicted hidden voxels. Each off-diagonal cell averages the two pairings of its models (original $i$ with the rebuilds of $j$, and original $j$ with the rebuilds of $i$); the outlined diagonal pairs each model with its own rebuilds. The diagonal averages 0.78, against 0.55 off the diagonal ($p < 0.001$, exact permutation test over all 40{,}320 ways of pairing the originals with the rebuilt models). The eight models come from one search and share many features, and iterations 238, 239 and 242 are at most four iterations apart, yet the diagonal is the largest value in its row for 4 of the 8 models and the largest or second-largest for 7 of them.
  \textbf{e}, \textbf{f}, Two of the eight descriptions, reproduced word for word. All eight descriptions, the rebuilt models, the agent logs and the evaluation code are \href{https://github.com/csinva/weight-space-programming-autoresearch}{on GitHub}.}
  \label{fig:story_bottleneck}
\end{figure}

\FloatBarrier

\subsection{ECoG responses to audio}
\label{sec:app_ecog}

\begin{figure}[!ht]
  \centering
  \includegraphics[width=\linewidth]{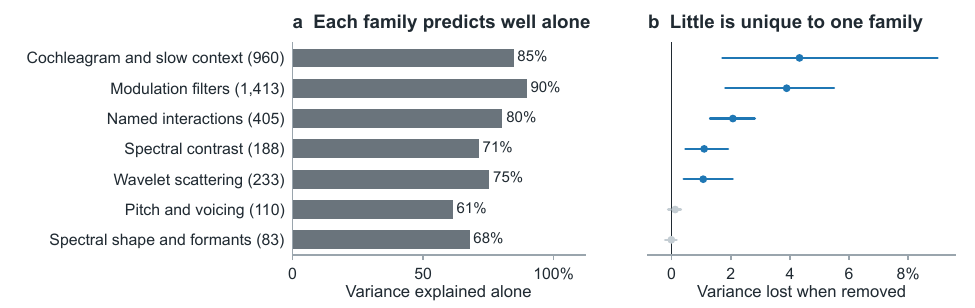}
  \caption{\textbf{Variance across families of written audio features.} We use all 166 electrodes and the final model. Reading its code revealed redundant columns and gates that never opened, so we removed them first (3{,}852 to 3{,}392 features, performance 0.0808 to 0.0805). The full model explains 1.58\% of the variance, measured as the mean signed squared correlation over electrodes. \textbf{a}, Variance explained by a ridge regression on one family alone, as a share of the full model, with feature counts in parentheses. \textbf{b}, Variance lost when we remove one family from the full model, again as a share of the full model. Bars show 95\% intervals from a paired bootstrap over the 23 visible segments, with intervals above zero in blue. These shares sum to 13\%, so the remaining 87\% is shared across families.}
  \label{fig:ecog_varpart}
\end{figure}

\begin{figure}[!ht]
  \centering
  \includegraphics[width=\linewidth]{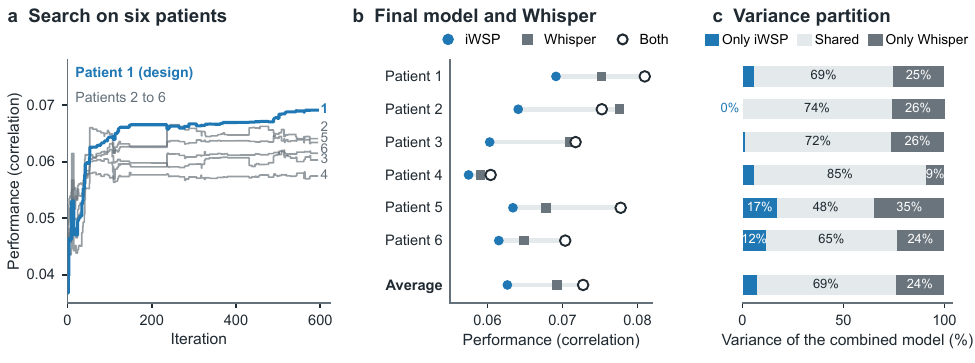}
  \caption{\textbf{The written ECoG features on six patients.} Every model of the search, designed on patient 1, is rescored on a hidden recording session of each patient, using the grid electrodes. \textbf{a}, Performance of the model the search kept (its best on patient 1's visible set so far) over iterations; numbers mark patients. \textbf{b}, The final iWSP model, Whisper, and both in one ridge regression. \textbf{c}, Variance of the combined model explained only by the iWSP features, by both, and only by Whisper, as a share of the combined model's explained variance.}
  \label{fig:ecog_multisubject}
\end{figure}

\paragraph{Transfer to five more patients.}
The written features were designed on one patient, so we test them on five more patients from the same dataset (\cref{fig:ecog_multisubject}). For each patient, we rescore every model of the search with the same features, delays and ridge regression, fit ridge regression on 500 segments (about 16 hours) and evaluate on the grid electrodes of one hidden recording session (107 to 416 segments, 3 to 13 hours). For the design patient, the hidden set is its 493 segments of the same session without the speech filter of the main text, because the other patients have no transcripts.
The early design transfers: on every patient, the models' performance tracks the design patient's (correlation 0.80 to 0.94 across models), and most of the gain comes within about the first 60 iterations, which build the cochleagram, compression and adaptation, and the modulation filters. The later refinements do not transfer: after iteration 170 they add 0.004 on the design patient but between $-0.002$ and 0.001 on the others. This suggests that some feature selection overfitting is possible in the limit when generalizing feature sets chosen by iWSP to other subjects. Whisper outperforms the iWSP model on every patient, by 0.001 to 0.013, yet the iWSP features explain variance that Whisper misses in four of the six patients (95\% interval above zero), up to 17\% of the combined model's variance. The modulation filters are the only family whose unique variance is positive on every patient (2\% to 7\% of the full model).

\begin{figure}[!ht]
  \centering
  \includegraphics[width=\linewidth]{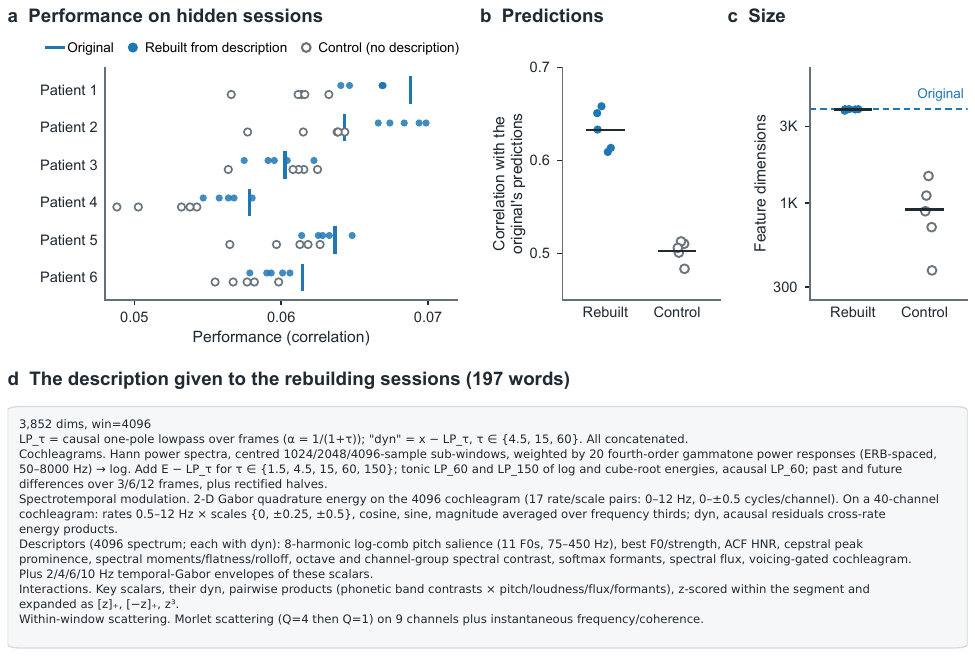}
  \caption{\textbf{A 197-word description is enough to rebuild the ECoG model.} A fresh session describes the final iWSP model in at most 200 words; five sessions rebuild it from the description, and five control sessions get the same task with no description, each once and without feedback. \textbf{a}, Performance on a hidden recording session of each patient, using the grid electrodes, for the original (vertical line), the five rebuilds (filled) and the five controls (open). \textbf{b}, Correlation between each model's hidden-set predictions and the original's, for patient 1; bars mark means. \textbf{c}, Feature dimensions (log scale); the dashed line marks the original's 3{,}852. \textbf{d}, The raw description.}
  \label{fig:ecog_bottleneck}
\end{figure}

\paragraph{A 200-word description is enough to rebuild the ECoG model.}
As for the other models (\cref{fig:loop}b), a fresh session with Claude Opus 5.5 reads the final model's code, with comments and run history removed, and describes it in at most 200 words (197 used; \cref{fig:ecog_bottleneck}d). Five fresh sessions rebuild the model from the description and the task alone, and five control sessions get the same task with no description. Every session runs once, in an isolated sandbox with no performance feedback, and we audit them as in \cref{sec:wsp}: no model fits anything to data, each produces identical features when the example audio changes, and no session reads files outside its sandbox.
On the hidden sessions of the six patients, the rebuilds stay within 0.003 of the original's performance, and on patient 2 they exceed it. They outperform the controls on five of the six patients and tie on the sixth (\cref{fig:ecog_bottleneck}a). The rebuilds' hidden-set predictions correlate with the original's at 0.63 on average, against 0.50 for the controls, and the rebuilds keep nearly all of the original's 3{,}852 dimensions (3{,}769 to 3{,}852), while the controls write models with 382 to 1{,}478 (\cref{fig:ecog_bottleneck}b, c).

\FloatBarrier

\subsection{Experiment costs}
\label{sec:app_costs}

\begin{table}[!ht]
  \centering
  \small
  \caption{\textbf{Agent cost of each experiment.} We give costs at API list prices, not counting the compute for running models and fitting ridge regressions. The exploratory rows for the algorithmic/NLP tasks and the story-listening fMRI searches estimate the earlier, non-final runs, which the paper does not report, at about double the cost of the final runs. The Haiku cost leaves out one run whose session was stopped before its cost was recorded. For the visual cortex experiments, the table lists the final search and the earlier exploratory runs separately; together they cost about \$\visTotalCost. The ECoG cost is a lower bound: it covers the June agent runs, including the precursor and clean-slate searches and runs stopped at their time limit, but not the interactive sessions that launched the agents.}
  \label{tab:costs}
  \input{tables/experiment_costs}
\end{table}
\FloatBarrier

\end{document}

%% file: tables/vision_numbers.tex
\newcommand{\visSearchModel}{Gemini 4 Carbon}
\newcommand{\visCandidates}{600}

\newcommand{\visSearchSubjects}{three}
\newcommand{\visFinalSubjects}{two}

\newcommand{\visNinetyIterations}{nine}

\newcommand{\visModelLines}{865}
\newcommand{\visLaionPooled}{0.466}
\newcommand{\visLaionEarly}{0.518}
\newcommand{\visLaionEarlyBest}{0.472}
\newcommand{\visLaionEarlyBestName}{Gabor filters with divisive normalization}
\newcommand{\visLaionEarlyGapMin}{0.044}
\newcommand{\visLaionEarlyGapMax}{0.047}
\newcommand{\visLaionVentral}{0.409}
\newcommand{\visLaionVentralBest}{0.309}
\newcommand{\visLaionVentralBestName}{Gabor filters with divisive normalization}
\newcommand{\visLaionVentralGapMin}{0.096}
\newcommand{\visLaionVentralGapMax}{0.105}
\newcommand{\visLaionImages}{747}
\newcommand{\visSearchSubjectsVisible}{0.536}
\newcommand{\visSearchSubjectsHidden}{0.536}
\newcommand{\visGaborNormSearch}{0.451}
\newcommand{\visGaborNormHeldOut}{0.395}
\newcommand{\visAlexEarlyLayer}{conv3}
\newcommand{\visAlexVentralLayer}{fc6}

\newcommand{\visNsdAlexVentral}{0.391}

\newcommand{\visNsdEarly}{0.443}
\newcommand{\visNsdEarlyBest}{0.400}

\newcommand{\visNsdEarlyGapMin}{0.035}
\newcommand{\visNsdEarlyGapMax}{0.054}
\newcommand{\visNsdVentral}{0.317}
\newcommand{\visNsdVentralBest}{0.256}

\newcommand{\visNsdVentralGapMin}{0.057}
\newcommand{\visNsdVentralGapMax}{0.064}
\newcommand{\visNsdImages}{998}
\newcommand{\visNsdSubjects}{four}
\newcommand{\visCombinedFeatures}{18{,}386}

\newcommand{\visNsdVentralCombined}{0.283}
\newcommand{\visRunCost}{539}
\newcommand{\visExploratoryCost}{4{,}500}
\newcommand{\visTotalCost}{5{,}000}

%% file: tables/wsp_task_list.tex
\begin{tabular}{@{}p{1.72in}p{2.35in}p{2.05in}@{}}
\toprule
Task & Definition & Example or format \\
\midrule
\multicolumn{3}{@{}l}{\textbf{Algorithmic, easier}} \\
\texttt{addition-five-digits} & Add two 5-digit non-negative integers. & \texttt{12345+67890=} $\to$ \texttt{080235} \\
\texttt{multiplication-five-digits} & Multiply two 5-digit non-negative integers. & \texttt{12345*67890=} $\to$ \texttt{0838102050} \\
\texttt{sort-five-digits} & Sort a string of 5 digits in ascending order. & \texttt{31415=} $\to$ \texttt{11345} \\
\texttt{digit-counting-10} & Count occurrences of a query digit in a string of 10 digits. & \texttt{31415926533=} $\to$ \texttt{02} \\
\texttt{parity-upto10-bits} & Compute parity (XOR) of a variable-length bit string, length 1-10. & \texttt{\_...1011010=} $\to$ \texttt{0} \\
\texttt{boolean-circuit-5-bits} & Evaluate a boolean expression of 5 bits joined by 4 binary operators. & \texttt{1\&0\textbar{}\allowbreak 1\textasciicircum{}1\&0=} $\to$ \texttt{0} \\
\texttt{linear-interpolation-two-points} & Linear interpolation given two integer (x, y) pairs and a query x. & \texttt{1,05;4,11;7=} $\to$ \texttt{17} \\
\midrule
\multicolumn{3}{@{}l}{\textbf{Algorithmic, harder}} \\
\texttt{word-reversal-3x3} & Reverse the order of 3 fixed-length words separated by underscores. & \texttt{cat\_dog\_pig=} $\to$ \texttt{pig\_dog\_cat} \\
\texttt{gcd-three-digits} & Greatest common divisor of two 3-digit non-negative integers. & \texttt{084,036=} $\to$ \texttt{012} \\
\texttt{decimal-to-binary-8bit} & Convert a decimal integer in [0, 255] to its 8-bit binary representation. & \texttt{042=} $\to$ \texttt{00101010} \\
\texttt{edit-distance-8} & Levenshtein edit distance between two 8-letter strings over \{a, b, c, d\}. & \texttt{abcdabcd,abddabca=} $\to$ \texttt{2} \\
\texttt{smallest-prime-factor} & Smallest prime factor of an integer below 100000 (a prime is its own smallest factor). & \texttt{03127=} $\to$ \texttt{00053} \\
\texttt{date-normalization} & Rewrite a date written in one of nine common formats as YYYY-MM-DD. & \texttt{march 3rd 2021\_...=} $\to$ \texttt{2021-03-03} \\
\midrule
\multicolumn{3}{@{}l}{\textbf{NLP}} \\
\texttt{trec-coarse} & Six-way question classification on a small subset of TREC (coarse labels). & \texttt{\textless{}question padded to 80 chars\textgreater{}=} $\to$ \texttt{L} \\
\texttt{sentiment-sst2} & Binary sentiment classification on a small subset of SST-2. & \texttt{\textless{}sentence padded to 80 chars\textgreater{}=} $\to$ \texttt{L} \\
\texttt{paraphrase-mrpc} & Binary paraphrase detection on a small subset of GLUE/MRPC. & \texttt{\textless{}s1 padded\textgreater{}\textbar{}\allowbreak \textless{}s2 padded\textgreater{}=} $\to$ \texttt{L} \\
\texttt{nli-snli} & Three-way natural language inference on a small subset of SNLI. & \texttt{\textless{}premise padded\textgreater{}\textbar{}\allowbreak \textless{}hypothesis padded\textgreater{}=} $\to$ \texttt{L} \\
\bottomrule
\end{tabular}

%% file: tables/wsp_model_comparison.tex
\begin{tabular}{@{}lrrr@{}}
\toprule
Measure & Opus & Sonnet & Haiku \\
\midrule
Iterations over 34 runs & 1,334 & 1,341 & 1,338 \\
Weights independent of task data & 1,303 & 1,042 & 1,311 \\
Weights depend on labels & 0 & 50 & 0 \\
Weights depend on inputs only & 0 & 34 & 0 \\
Could not be checked & 31 & 215 & 27 \\
Algorithmic runs at 100\% visible accuracy & 26 of 26 & 26 of 26 & 25 of 26 \\
Algorithmic runs at 100\% hidden accuracy & 26 of 26 & 25 of 26 & 25 of 26 \\
Median smallest model at 100\% visible accuracy (parameters) & 166 & 41,600 & 100,000 \\
TREC hidden accuracy & 87.3\% & 86.6\% & 71.1\% \\
\bottomrule
\end{tabular}

%% file: tables/wsp_bottleneck_descriptions.tex
Word reversal (188) & This model has no MLP: one causal attention read acts as a pointer, followed by a nearest-value lookup. Attention: one layer, one head, using hard argmax instead of softmax. Each position's key is its 1-based position. The query at each generation step is the slot to copy. With i = position $-$ 11 (the "=" is position 11) and word = floor(i/4), the query is 9 + i $-$ 8$\cdot$word. The score is negative squared distance between query and key, and the single best earlier position wins. So the answer copies prompt indices 8, 9, 10, 11, 4, 5, 6, 7, 0, 1, 2: word 3, separator, word 2, separator, word 1. Value and readout: the value is the token id as one scalar, capped at the separator's id, so "=" reads as "\_". That lets the first separator be copied from the "=" slot. Logits are the negative squared difference between the read value and each vocabulary id. An exact match scores 0; ties go to the lower id, "\_". Errors: selection is exact with no softmax leakage, so accuracy should be 100\%, even with repeated letters. \\[3pt]
Boolean circuit (186) & A tiny recurrent "cellular automaton" transformer. Each position holds one scalar state, the expression value so far, stored centred as S $-$ 0.5. One tied block (one attention head, 2-neuron ReLU MLP) is applied 10 times; each pass moves information one cell right, enough to reach '='. Attention has no key matrix: positions are sinusoidal angles at 30$^\circ$ per step, and the query is the position rotated back one step, scaled by 200, giving near-hard attention to the previous cell (position 0 attends to itself). Values copy that cell's AND/XOR flags and put its state on the bit channel; each cell adds its own embedding. Op tokens carry their flag plus identity bit (\& 1, $|$ and \^{} 0), so op cells and '=' (zero embedding) pass the state through. The MLP does one fold step: new state = relu(b + S $-$ A) $-$ relu(2(b + S) + X $-$ A $-$ 2), giving OR with no flags, AND with A, XOR with X. State starts at 0; readout logits are $\pm$4 times the centred state. Errors: expected none (attention leakage $\approx$ e\^{}$-$27), though untested. \\[3pt]
Multiplication (190) & This model has no MLP and no LayerNorm: two attention-only layers on a 7-channel float64 residual stream whose registers hold integers up to about 1e10. Layer 1 (one head) makes each B digit compute b$\cdot$A. A query at a B digit b scores multiplicand position i at log(b$\cdot$10\^{}(4$-$i)) against a huge zero-valued sink key at '$*$'. Because the sink dominates (scale about 1e17), softmax acts like a plain exponential, so the head reads b$\cdot$A/scale, rescaled to b$\cdot$A. Digit 0 attends only to the sink. In layer 2, head 1 weights B positions proportionally to 10\^{}(10$-$t); multiplying the average by 11111 gives P = A$\cdot$B. Head 2 uses the same sink trick (scale about 1e22) to read V, the value of already-emitted answer digits times their place values, and subtracts it, so X = P $-$ V + 0.5, a long-division remainder. The readout is a tropical decoder: logit\_k = k$\cdot$X $-$ place$\cdot$k(k+1)/2 with place = 10\^{}(20$-$t) at answer position t, whose argmax is floor(X/place). Expect essentially 100\% exact match; leakage is about 1e-12 relative, so float64 is required (float32 breaks). One wrong digit corrupts the remainder and all later digits. \\[3pt]
GCD (189) & Tokens embed as one scalar (digit value; comma and "=" are 0) plus a fixed one-hot position code. The residual state has channels VAL, A, B, X. A six-head parser attention layer uses huge logits (about 40), so each head reads one operand digit, scaled by its place value (100, 10, 1). This gives exact integers A and B at answer positions. The comma attends to itself, keeping A=B=0 there as a "zero" slot. Euclid runs as residual ReLU MLPs on (A, B): 12 tied rounds of four folds with scales 64, 16, 4, 1. Each fold maps A to its distance from the nearest multiple of s$\cdot$B. This triangular wave uses 4 ReLU knots at m$\cdot$s$\cdot$B/2 with alternating $\pm$2 slopes and is valid only while A $\leq$ 2sB, hence radix 4. The s=1 fold also swaps (A,B)$\rightarrow$(B, fold). The result settles into a 2-cycle between (g,0) and (0,g), so A+B=g. Readout attention forms X=(A+B)/100, /10 or /1 by mixing with the zero slot, minus 100$\times$ and 10$\times$ earlier emitted digits. The head scores digit d as $-$(X$-$d$-$0.495)$^2$, flooring X. The file claims zero errors on all 10$^6$ pairs. \\[3pt]
Smallest prime factor (181) & Two-block transformer, no LayerNorm, float64, all signals scalar (not one-hot). It should be exactly correct for every n from 2 to 99999, so there is no error pattern to match; aim for 100\%. Block 1 attention: one head, constant query, keys (4$-$i)$\cdot$ln10 for prompt digit i, $-$60 elsewhere. Softmax weights become 10\^{}(4$-$i)/11111, so summing digit values times 11111 yields n. Block 1 MLP (sieve): one neuron per prime up to 316 (65 primes) with input n/p and pulse activation relu(1 + 10$^4$(cos2$\pi$u $-$ 1)), firing only at integers; non-divisors are at least 1/313 away. Lateral inhibition (2$\cdot$pulse minus cumulative sum over smaller primes) keeps only the smallest firing prime, which writes its three digits. Block 2 attention: Gaussian copy head, score $-$40(i$-$(s$-$5))$^2$, so output position s reads prompt digit s$-$5. Block 2 MLP: ReLU gate. Any nonzero factor digit suppresses the copy (weight $-$9); flags on the last three output steps release the factor's hundreds, tens and units digits. No factor means n is prime and gets copied. Readout: logit\_d = 2d$\cdot$A $-$ d$^2$ picks the nearest digit; "=" is blocked. \\[3pt]
TREC (196) & A hand-built keyword decision list with no training. Block 1: each character is embedded as a 5-bit letter code with flags for word boundary, start and padding, plus position. Head h of 20 copies the character h steps back using a Gaussian position kernel. Each MLP neuron is an exact AND gate for one of about 400 keyword patterns: words, stems, or question-initial words. Block 2: '=' attends to the highest salience, 300 $\times$ priority minus 3 $\times$ position, so the earliest match wins ties. It copies that position's class vote. If nothing matches, the answer is ENTY. Priority order: ABBR for "stands for" and "abbreviat"; DESC for "comes from"; then initial "who" (HUM), "where" (LOC), initial "when" (NUM) and "why" (DESC); definition or cause words such as mean, origin and reason (DESC); "how did/does/is" (DESC); other "how" (NUM); "what is X" questions under 17 characters (DESC); topical nouns, with people HUM, places LOC, measures NUM and things ENTY; "called" and "what kind" (ENTY); "what are the" (ENTY); and last, initial "what is" (DESC). Errors come from brittle surface matching: missing keywords, the wrong noun winning, and long "what is the" questions falling through to DESC. \\[3pt]

%% file: tables/fmri_unique_variance.tex
\begin{tabular}{lrrrrrr}
\toprule
Computation & V1 & V2 & V3 & hV4 & FFA & PPA \\
\midrule
Gabor energy, local mean & $2.0$ {\scriptsize$\pm$0.6} & $1.8$ {\scriptsize$\pm$0.5} & $1.4$ {\scriptsize$\pm$0.5} & $0.2$ {\scriptsize$\pm$0.3} & $-1.0$ {\scriptsize$\pm$0.2} & $-0.4$ {\scriptsize$\pm$0.1} \\
Gabor energy, local variability & $0.2$ {\scriptsize$\pm$0.1} & $0.4$ {\scriptsize$\pm$0.1} & $0.5$ {\scriptsize$\pm$0.1} & $0.7$ {\scriptsize$\pm$0.2} & $0.4$ {\scriptsize$\pm$0.2} & $1.1$ {\scriptsize$\pm$0.2} \\
Blobs and luminance & $0.2$ {\scriptsize$\pm$0.0} & $0.1$ {\scriptsize$\pm$0.0} & $0.0$ {\scriptsize$\pm$0.0} & $-0.1$ {\scriptsize$\pm$0.1} & $-0.2$ {\scriptsize$\pm$0.1} & $-0.3$ {\scriptsize$\pm$0.1} \\
Color contrast & $0.5$ {\scriptsize$\pm$0.1} & $0.5$ {\scriptsize$\pm$0.1} & $0.4$ {\scriptsize$\pm$0.2} & $0.3$ {\scriptsize$\pm$0.2} & $-0.5$ {\scriptsize$\pm$0.7} & $0.0$ {\scriptsize$\pm$0.2} \\
Surface geometry, local mean & $1.5$ {\scriptsize$\pm$0.1} & $1.4$ {\scriptsize$\pm$0.1} & $1.4$ {\scriptsize$\pm$0.2} & $1.3$ {\scriptsize$\pm$0.1} & $1.3$ {\scriptsize$\pm$0.2} & $0.7$ {\scriptsize$\pm$0.1} \\
Surface geometry, local variability & $-0.3$ {\scriptsize$\pm$0.1} & $-0.1$ {\scriptsize$\pm$0.2} & $0.3$ {\scriptsize$\pm$0.2} & $1.0$ {\scriptsize$\pm$0.1} & $0.8$ {\scriptsize$\pm$0.3} & $-0.8$ {\scriptsize$\pm$0.5} \\
Contour curvature & $0.0$ {\scriptsize$\pm$0.1} & $-0.1$ {\scriptsize$\pm$0.0} & $-0.2$ {\scriptsize$\pm$0.1} & $-0.3$ {\scriptsize$\pm$0.1} & $0.8$ {\scriptsize$\pm$0.5} & $0.6$ {\scriptsize$\pm$0.2} \\
Line versus edge phase & $0.1$ {\scriptsize$\pm$0.1} & $0.0$ {\scriptsize$\pm$0.1} & $-0.1$ {\scriptsize$\pm$0.1} & $-0.1$ {\scriptsize$\pm$0.2} & $-0.5$ {\scriptsize$\pm$0.3} & $0.1$ {\scriptsize$\pm$0.0} \\
Second-order filtering & $-0.1$ {\scriptsize$\pm$0.0} & $-0.1$ {\scriptsize$\pm$0.0} & $0.0$ {\scriptsize$\pm$0.0} & $0.1$ {\scriptsize$\pm$0.1} & $0.2$ {\scriptsize$\pm$0.1} & $-0.1$ {\scriptsize$\pm$0.1} \\
Channel correlations & $-0.1$ {\scriptsize$\pm$0.1} & $-0.1$ {\scriptsize$\pm$0.1} & $0.0$ {\scriptsize$\pm$0.1} & $0.7$ {\scriptsize$\pm$0.2} & $4.9$ {\scriptsize$\pm$0.6} & $1.2$ {\scriptsize$\pm$0.2} \\
Cross-scale relations & $-0.2$ {\scriptsize$\pm$0.0} & $-0.1$ {\scriptsize$\pm$0.0} & $-0.1$ {\scriptsize$\pm$0.0} & $0.0$ {\scriptsize$\pm$0.1} & $0.8$ {\scriptsize$\pm$0.2} & $0.9$ {\scriptsize$\pm$0.3} \\
Geometry correlations & $0.0$ {\scriptsize$\pm$0.1} & $0.0$ {\scriptsize$\pm$0.1} & $0.1$ {\scriptsize$\pm$0.1} & $0.6$ {\scriptsize$\pm$0.2} & $3.0$ {\scriptsize$\pm$0.3} & $1.7$ {\scriptsize$\pm$0.3} \\
Color layout & $0.4$ {\scriptsize$\pm$0.1} & $0.4$ {\scriptsize$\pm$0.1} & $0.4$ {\scriptsize$\pm$0.1} & $0.3$ {\scriptsize$\pm$0.1} & $0.9$ {\scriptsize$\pm$0.2} & $1.5$ {\scriptsize$\pm$0.1} \\
Symmetry and blob count & $-0.1$ {\scriptsize$\pm$0.1} & $0.0$ {\scriptsize$\pm$0.0} & $0.2$ {\scriptsize$\pm$0.1} & $0.2$ {\scriptsize$\pm$0.2} & $0.8$ {\scriptsize$\pm$0.2} & $0.2$ {\scriptsize$\pm$0.1} \\
\midrule
Not unique to one computation & $96.0$ {\scriptsize$\pm$0.3} & $95.9$ {\scriptsize$\pm$0.1} & $95.6$ {\scriptsize$\pm$0.1} & $95.2$ {\scriptsize$\pm$0.3} & $88.2$ {\scriptsize$\pm$0.7} & $93.5$ {\scriptsize$\pm$0.5} \\
\bottomrule
\end{tabular}

%% file: tables/fmri_bottleneck_complexity.tex
\begin{tabular}{@{}lrrrr@{}}
\toprule
Model & Lines of code & Numeric constants & Output features & Pooled $r$ \\
\midrule
Original & 742 & 1{,}004 & 9{,}996 & 0.488 \\
Rebuild 1 & 355 & 297 & 9{,}206 & 0.447 \\
Rebuild 2 & 292 & 311 & 9{,}398 & 0.452 \\
Rebuild 3 & 358 & 317 & 9{,}165 & 0.454 \\
Rebuild 4 & 347 & 343 & 9{,}530 & 0.452 \\
\midrule
Rebuild average & 338 & 317 & 9{,}325 & 0.451 \\
\bottomrule
\end{tabular}

%% file: tables/story_hidden_set.tex
\begin{tabular}{@{}llrrr@{}}
\toprule
Iteration & Model & Visible & Hidden & Hidden, edges removed \\
\midrule
 & GPT-2 XL & 0.079 & 0.046 & 0.047 \\
1 & \texttt{BoCharMean-d256} & 0.031 & 0.016 & 0.017 \\
40 & \texttt{MultiScaleSubword+FuncWordTags} & 0.040 & 0.020 & 0.021 \\
76 & \texttt{AllNewFeats-Words15} & 0.044 & 0.022 & 0.023 \\
152 & \texttt{ExpDec-R4.0} & 0.044 & 0.022 & 0.023 \\
192 & \texttt{DenseSemXX-N20} & 0.056 & 0.026 & 0.027 \\
238 (before jump) & \texttt{WordNetMorphLingMultiTau} & 0.067 & 0.032 & 0.030 \\
239 (after jump) & \texttt{WordNetMorphLingStoryArc} & 0.104 & 0.043 & 0.023 \\
242 (final) & \texttt{WordNetMorphLingNovelty} & 0.114 & 0.048 & 0.026 \\
\bottomrule
\end{tabular}

%% file: tables/experiment_costs.tex
\begin{tabular}{@{}lr@{}}
\toprule
Experiment & Cost at API list price \\
\midrule
Algorithmic/NLP tasks, Claude Opus 5.5 (34 runs) & \$449 \\
Algorithmic/NLP tasks, Claude Sonnet 5.5 (34 runs) & \$102 \\
Algorithmic/NLP tasks, Claude Haiku 5.5 (34 runs) & \$71 \\
Algorithmic/NLP tasks, exploratory runs & about \$1,240 (estimate) \\
Rebuilding algorithmic/NLP models from descriptions, Claude Opus 5.5 & \$60 \\
Rebuilding language fMRI models from descriptions, Claude Opus 5.5 & \$10 \\
Visual cortex search, \visSearchModel{} & \$\visRunCost{} \\
Visual cortex exploratory runs & about \$\visExploratoryCost{} \\
Story-listening fMRI searches, Claude Opus 4.7 & about \$3,000 (estimate) \\
Story-listening fMRI exploratory runs & about \$6,000 (estimate) \\
ECoG searches, Claude Opus 4.7 & about \$1,540 \\
\midrule
Total & about \$17,500 \\
\bottomrule
\end{tabular}

%% file: refs.bib
@article{naselaris2011encoding,
  title={Encoding and decoding in {fMRI}},
  author={Naselaris, Thomas and Kay, Kendrick N and Nishimoto, Shinji and Gallant, Jack L},
  journal={NeuroImage},
  volume={56},
  number={2},
  pages={400--410},
  year={2011}
}

@article{goldstein2023deep,
  title={Deep speech-to-text models capture the neural basis of spontaneous speech in everyday conversations},
  author={Goldstein, Ariel and Wang, Haocheng and Niekerken, Leonard and Zada, Zaid and Aubrey, Bobbi and Sheffer, Tom and Nastase, Samuel A and Gazula, Harshvardhan and Schain, Mariano and Singh, Aditi and others},
  journal={bioRxiv},
  pages={2023--06},
  year={2023},
  publisher={Cold Spring Harbor Laboratory}
}

@article{singh2025evaluating,
  title={Evaluating scientific theories as predictive models in language neuroscience},
  author={Singh, Chandan and Antonello, Richard J and Guo, Sihang and Mischler, Gavin and Gao, Jianfeng and Mesgarani, Nima and Huth, Alexander G},
  journal={BioRxiv},
  year={2025}
}

@article{yamins2014performance,
  title={Performance-optimized hierarchical models predict neural responses in higher visual cortex},
  author={Yamins, Daniel L K and Hong, Ha and Cadieu, Charles F and Solomon, Ethan A and Seibert, Darren and DiCarlo, James J},
  journal={Proceedings of the National Academy of Sciences},
  volume={111},
  number={23},
  pages={8619--8624},
  year={2014}
}

@article{khaligh2014deep,
  title={Deep supervised, but not unsupervised, models may explain {IT} cortical representation},
  author={Khaligh-Razavi, Seyed-Mahdi and Kriegeskorte, Nikolaus},
  journal={PLoS Computational Biology},
  volume={10},
  number={11},
  pages={e1003915},
  year={2014}
}

@article{cadena2019deep,
  title={Deep convolutional models improve predictions of macaque {V1} responses to natural images},
  author={Cadena, Santiago A and Denfield, George H and Walker, Edgar Y and Gatys, Leon A and Tolias, Andreas S and Bethge, Matthias and Ecker, Alexander S},
  journal={PLoS Computational Biology},
  volume={15},
  number={4},
  pages={e1006897},
  year={2019}
}

@article{zhuang2021unsupervised,
  title={Unsupervised neural network models of the ventral visual stream},
  author={Zhuang, Chengxu and Yan, Siming and Nayebi, Aran and Schrimpf, Martin and Frank, Michael C and DiCarlo, James J and Yamins, Daniel L K},
  journal={Proceedings of the National Academy of Sciences},
  volume={118},
  number={3},
  pages={e2014196118},
  year={2021}
}

@article{konkle2022self,
  title={A self-supervised domain-general learning framework for human ventral stream representation},
  author={Konkle, Talia and Alvarez, George A},
  journal={Nature Communications},
  volume={13},
  pages={491},
  year={2022}
}

@article{schrimpf2018brain,
  title={Brain-{S}core: which artificial neural network for object recognition is most brain-like?},
  author={Schrimpf, Martin and Kubilius, Jonas and Hong, Ha and Majaj, Najib J and Rajalingham, Rishi and Issa, Elias B and Kar, Kohitij and Bashivan, Pouya and Prescott-Roy, Jonathan and Geiger, Franziska and others},
  journal={bioRxiv},
  year={2018},
  doi={10.1101/407007}
}

@article{adelson1985spatiotemporal,
  title={Spatiotemporal energy models for the perception of motion},
  author={Adelson, Edward H and Bergen, James R},
  journal={Journal of the Optical Society of America A},
  volume={2},
  number={2},
  pages={284--299},
  year={1985}
}

@article{riesenhuber1999hierarchical,
  title={Hierarchical models of object recognition in cortex},
  author={Riesenhuber, Maximilian and Poggio, Tomaso},
  journal={Nature Neuroscience},
  volume={2},
  pages={1019--1025},
  year={1999}
}

@article{chen2025universal,
  title={Universal dimensions of visual representation},
  author={Chen, Zirui and Bonner, Michael F},
  journal={Science Advances},
  volume={11},
  number={27},
  pages={eadw7697},
  year={2025}
}

@inproceedings{krizhevsky2012imagenet,
  title={ImageNet classification with deep convolutional neural networks},
  author={Krizhevsky, Alex and Sutskever, Ilya and Hinton, Geoffrey E},
  booktitle={Advances in Neural Information Processing Systems},
  volume={25},
  year={2012}
}

@article{novikov2025alphaevolve,
  title={Alphaevolve: A coding agent for scientific and algorithmic discovery},
  author={Novikov, Alexander and V{\~u}, Ng{\^a}n and Eisenberger, Marvin and Dupont, Emilien and Huang, Po-Sen and Wagner, Adam Zsolt and Shirobokov, Sergey and Kozlovskii, Borislav and Ruiz, Francisco JR and Mehrabian, Abbas and others},
  journal={arXiv preprint arXiv:2506.13131},
  year={2025}
}

@article{lu2024ai,
  title={The ai scientist: Towards fully automated open-ended scientific discovery},
  author={Lu, Chris and Lu, Cong and Lange, Robert Tjarko and Foerster, Jakob and Clune, Jeff and Ha, David},
  journal={arXiv preprint arXiv:2408.06292},
  year={2024}
}

@article{chen2024scienceagentbench,
  title={Scienceagentbench: Toward rigorous assessment of language agents for data-driven scientific discovery},
  author={Chen, Ziru and Chen, Shijie and Ning, Yuting and Zhang, Qianheng and Wang, Boshi and Yu, Botao and Li, Yifei and Liao, Zeyi and Wei, Chen and Lu, Zitong and others},
  journal={arXiv preprint arXiv:2410.05080},
  year={2024}
}

@article{liu2024infini,
  title={Infini-gram: Scaling unbounded n-gram language models to a trillion tokens},
  author={Liu, Jiacheng and Min, Sewon and Zettlemoyer, Luke and Choi, Yejin and Hajishirzi, Hannaneh},
  journal={arXiv preprint arXiv:2401.17377},
  year={2024}
}

@article{benara2024crafting,
  title={Crafting Interpretable Embeddings by Asking LLMs Questions},
  author={Benara, Vinamra and Singh, Chandan and Morris, John X and Antonello, Richard and Stoica, Ion and Huth, Alexander G and Gao, Jianfeng},
  journal={arXiv preprint arXiv:2405.16714},
  year={2024}
}

@article{ifargan2024autonomous,
  title={Autonomous LLM-driven research from data to human-verifiable research papers},
  author={Ifargan, Tal and Hafner, Lukas and Kern, Maor and Alcalay, Ori and Kishony, Roy},
  journal={arXiv preprint arXiv:2404.17605},
  year={2024}
}

@misc{wu2024retrieval,
      title={Retrieval Head Mechanistically Explains Long-Context Factuality}, 
      author={Wenhao Wu and Yizhong Wang and Guangxuan Xiao and Hao Peng and Yao Fu},
      year={2024},
      eprint={2404.15574},
      archivePrefix={arXiv},
      primaryClass={cs.CL}
}

@article{merullo2023circuit,
  title={Circuit Component Reuse Across Tasks in Transformer Language Models},
  author={Merullo, Jack and Eickhoff, Carsten and Pavlick, Ellie},
  journal={arXiv preprint arXiv:2310.08744},
  year={2023}
}

@article{gabriel2020artificial,
  title={Artificial intelligence, values, and alignment},
  author={Gabriel, Iason},
  journal={Minds and machines},
  volume={30},
  number={3},
  pages={411--437},
  year={2020},
  publisher={Springer}
}

@article{huth2016natural,
  title={Natural speech reveals the semantic maps that tile human cerebral cortex},
  author={Huth, Alexander G and De Heer, Wendy A and Griffiths, Thomas L and Theunissen, Fr{\'e}d{\'e}ric E and Gallant, Jack L},
  journal={Nature},
  volume={532},
  number={7600},
  pages={453--458},
  year={2016},
  publisher={Nature Publishing Group}
}

@article{amodei2016concrete,
  title={Concrete problems in {AI} safety},
  author={Amodei, Dario and Olah, Chris and Steinhardt, Jacob and Christiano, Paul and Schulman, John and Man{\'e}, Dan},
  journal={arXiv preprint arXiv:1606.06565},
  year={2016}
}

@article{romera2023mathematical,
  title={Mathematical discoveries from program search with large language models},
  author={Romera-Paredes, Bernardino and Barekatain, Mohammadamin and Novikov, Alexander and Balog, Matej and Kumar, M Pawan and Dupont, Emilien and Ruiz, Francisco JR and Ellenberg, Jordan S and Wang, Pengming and Fawzi, Omar and others},
  journal={Nature},
  pages={1--3},
  year={2023},
  publisher={Nature Publishing Group UK London}
}

@inproceedings{hernandez2022natural,
  title={Natural language descriptions of deep visual features},
  author={Hernandez, Evan and Schwettmann, Sarah and Bau, David and Bagashvili, Teona and Torralba, Antonio and Andreas, Jacob},
  booktitle={International Conference on Learning Representations},
  year={2022}
}

@article{lebel2022natural,
  title={A natural language fMRI dataset for voxelwise encoding models},
  author={LeBel, Amanda and Wagner, Lauren and Jain, Shailee and Adhikari-Desai, Aneesh and Gupta, Bhavin and Morgenthal, Allyson and Tang, Jerry and Xu, Lixiang and Huth, Alexander G},
  journal={bioRxiv},
  pages={2022--09},
  year={2022},
  publisher={Cold Spring Harbor Laboratory}
}

@article{schrimpf2021neural,
  title={The neural architecture of language: Integrative modeling converges on predictive processing},
  author={Schrimpf, Martin and Blank, Idan Asher and Tuckute, Greta and Kauf, Carina and Hosseini, Eghbal A and Kanwisher, Nancy and Tenenbaum, Joshua B and Fedorenko, Evelina},
  journal={Proceedings of the National Academy of Sciences},
  volume={118},
  number={45},
  pages={e2105646118},
  year={2021},
  publisher={National Acad Sciences}
}

@inproceedings{shaham2024multimodal,
  title={A multimodal automated interpretability agent},
  author={Shaham, Tamar Rott and Schwettmann, Sarah and Wang, Franklin and Rajaram, Achyuta and Hernandez, Evan and Andreas, Jacob and Torralba, Antonio},
  booktitle={Forty-first International Conference on Machine Learning},
  year={2024}
}

@misc{bills2023language,
  title={Language models can explain neurons in language models},
  author={Bills, Steven and Cammarata, Nick and Mossing, Dan and Tillman, Henk and Gao, Leo and Goh, Gabriel and Sutskever, Ilya and Leike, Jan and Wu, Jeff and Saunders, William},
  url={https://openaipublic.blob.core.windows.net/neuron-explainer/paper/index. html},
  comment = {(Date accessed: 14.05. 2023)},
  year={2023}
}

@article{radford2019language,
  title={Language models are unsupervised multitask learners},
  author={Radford, Alec and Wu, Jeffrey and Child, Rewon and Luan, David and Amodei, Dario and Sutskever, Ilya and others},
  journal={OpenAI blog},
  volume={1},
  number={8},
  pages={9},
  year={2019}
}

@article{jain2018incorporating,
  title={Incorporating context into language encoding models for fMRI},
  author={Jain, Shailee and Huth, Alexander},
  journal={Advances in neural information processing systems},
  volume={31},
  year={2018}
}

@article{joulin2016bag,
  title={Bag of tricks for efficient text classification},
  author={Joulin, Armand and Grave, Edouard and Bojanowski, Piotr and Mikolov, Tomas},
  journal={arXiv preprint arXiv:1607.01759},
  year={2016}
}

@article{meng2022locating,
  title={Locating and editing factual knowledge in {GPT}},
  author={Meng, Kevin and Bau, David and Andonian, Alex and Belinkov, Yonatan},
  journal={arXiv preprint arXiv:2202.05262},
  year={2022}
}

@article{elhage2021mathematical,
  title={A mathematical framework for transformer circuits},
  author={Elhage, Nelson and Nanda, Neel and Olsson, Catherine and Henighan, Tom and Joseph, Nicholas and Mann, Ben and Askell, Amanda and Bai, Yuntao and Chen, Anna and Conerly, Tom and others},
  journal={Transformer Circuits Thread},
  volume={1},
  year={2021}
}

@inproceedings{koh2020concept,
  title={Concept bottleneck models},
  author={Koh, Pang Wei and Nguyen, Thao and Tang, Yew Siang and Mussmann, Stephen and Pierson, Emma and Kim, Been and Liang, Percy},
  booktitle={International Conference on Machine Learning},
  pages={5338--5348},
  year={2020},
  organization={PMLR}
}

@inproceedings{caruana2015intelligible,
  title={Intelligible models for healthcare: Predicting pneumonia risk and hospital 30-day readmission},
  author={Caruana, Rich and Lou, Yin and Gehrke, Johannes and Koch, Paul and Sturm, Marc and Elhadad, Noemie},
  booktitle={Proceedings of the 21th ACM SIGKDD international conference on knowledge discovery and data mining},
  pages={1721--1730},
  year={2015}
}

@article{bruna2013invariant,
	title        = {Invariant scattering convolution networks},
	author       = {Bruna, Joan and Mallat, St{\'e}phane},
	year         = 2013,
	journal      = {IEEE transactions on pattern analysis and machine intelligence},
	publisher    = {IEEE},
	volume       = 35,
	number       = 8,
	pages        = {1872--1886}
}

@inproceedings{lundberg2017unified,
	title        = {A unified approach to interpreting model predictions},
	author       = {Lundberg, Scott M and Lee, Su-In},
	year         = 2017,
	booktitle    = {Advances in Neural Information Processing Systems},
	pages        = {4768--4777}
}

@inproceedings{ribeiro2016should,
	title        = {Why should {I} trust you?: Explaining the predictions of any classifier},
	author       = {Ribeiro, Marco Tulio and Singh, Sameer and Guestrin, Carlos},
	year         = 2016,
	booktitle    = {Proceedings of the 22nd ACM SIGKDD International Conference on Knowledge Discovery and Data Mining},
	pages        = {1135--1144},
	organization = {ACM}
}

@article{ross2017right,
	title        = {Right for the right reasons: Training differentiable models by constraining their explanations},
	author       = {Ross, Andrew Slavin and Hughes, Michael C and Doshi-Velez, Finale},
	year         = 2017,
	journal      = {arXiv preprint arXiv:1703.03717}
}

@article{schramowski2020making,
	title        = {Making deep neural networks right for the right scientific reasons by interacting with their explanations},
	author       = {Schramowski, Patrick and Stammer, Wolfgang and Teso, Stefano and Brugger, Anna and Herbert, Franziska and Shao, Xiaoting and Luigs, Hans-Georg and Mahlein, Anne-Katrin and Kersting, Kristian},
	year         = 2020,
	journal      = {Nature Machine Intelligence},
	publisher    = {Nature Publishing Group},
	volume       = 2,
	number       = 8,
	pages        = {476--486}
}

@inproceedings{selvaraju2017grad,
	title        = {Grad-cam: Visual explanations from deep networks via gradient-based localization},
	author       = {Selvaraju, Ramprasaath R and Cogswell, Michael and Das, Abhishek and Vedantam, Ramakrishna and Parikh, Devi and Batra, Dhruv},
	year         = 2017,
	booktitle    = {Proceedings of the IEEE international conference on computer vision},
	pages        = {618--626}
}

@article{singh2019Hierarchical,
	title        = {Hierarchical Interpretations for Neural Network Predictions},
	author       = {Singh, Chandan and Murdoch, W James and Yu, Bin},
	year         = 2019,
	journal      = {International Conference on Learning Representations},
	booktitle    = {International Conference on Learning Representations},
	pages        = 26,
	url          = {https://openreview.net/forum?id=SkEqro0ctQ},
	language     = {en},
	date         = 2019
}

@article{singh2021imodels,
	title        = {imodels: a python package for fitting interpretable models},
	author       = {Singh, Chandan and Nasseri, Keyan and Tan, Yan Shuo and Tang, Tiffany and Yu, Bin},
	year         = 2021,
	journal      = {Journal of Open Source Software},
	publisher    = {The Open Journal},
	volume       = 6,
	number       = 61,
	pages        = 3192,
	xdoi          = {10.21105/joss.03192},
	xurl          = {https://doi.org/10.21105/joss.03192}
}

@article{singh2023explainingmodules,
  title={Explaining black box text modules in natural language with language models},
  author={Singh, Chandan and Hsu, Aliyah R and Antonello, Richard and Jain, Shailee and Huth, Alexander G and Yu, Bin and Gao, Jianfeng},
  journal={arXiv preprint arXiv:2305.09863},
  year={2023}
}

@article{singh2023augmenting,
  title={Augmenting interpretable models with large language models during training},
  author={Singh, Chandan and Askari, Armin and Caruana, Rich and Gao, Jianfeng},
  journal={Nature Communications},
  volume={14},
  number={1},
  pages={7913},
  year={2023},
  publisher={Nature Publishing Group UK London}
}

@article{feng2026human,
  title={Human-AI Co-design for Clinical Prediction Models},
  author={Feng, Jean and Kothari, Avni and Vossler, Patrick and Bishara, Andrew and Zier, Lucas and Addo, Newton and Kornblith, Aaron and Tan, Yan Shuo and Singh, Chandan},
  journal={arXiv preprint arXiv:2601.09072},
  year={2026}
}

@article{olsson2022context,
  title={In-context learning and induction heads},
  author={Olsson, Catherine and Elhage, Nelson and Nanda, Neel and Joseph, Nicholas and DasSarma, Nova and Henighan, Tom and Mann, Ben and Askell, Amanda and Bai, Yuntao and Chen, Anna and others},
  journal={arXiv preprint arXiv:2209.11895},
  year={2022}
}

@article{lieberum2023does,
  title={Does circuit analysis interpretability scale? {E}vidence from multiple choice capabilities in chinchilla},
  author={Lieberum, Tom and Rahtz, Matthew and Kram{\'a}r, J{\'a}nos and Irving, Geoffrey and Shah, Rohin and Mikulik, Vladimir},
  journal={arXiv preprint arXiv:2307.09458},
  year={2023}
}

@article{wang2022interpretability,
  title={Interpretability in the wild: a circuit for indirect object identification in {GPT}-2 small},
  author={Wang, Kevin and Variengien, Alexandre and Conmy, Arthur and Shlegeris, Buck and Steinhardt, Jacob},
  journal={arXiv preprint arXiv:2211.00593},
  year={2022}
}

@article{swanson2025virtual,
  title={The Virtual Lab of AI agents designs new SARS-CoV-2 nanobodies},
  author={Swanson, Kyle and Wu, Wesley and Bulaong, Nash L and Pak, John E and Zou, James},
  journal={Nature},
  volume={646},
  number={8085},
  pages={716--723},
  year={2025},
  publisher={Nature Publishing Group UK London}
}

@article{luo2025more,
  title={The More You Automate, the Less You See: Hidden Pitfalls of AI Scientist Systems},
  author={Luo, Ziming and Kasirzadeh, Atoosa and Shah, Nihar B},
  journal={arXiv preprint arXiv:2509.08713},
  year={2025}
}

@article{gottweis2025towards,
  title={Towards an AI co-scientist},
  author={Gottweis, Juraj and Weng, Wei-Hung and Daryin, Alexander and Tu, Tao and Palepu, Anil and Sirkovic, Petar and Myaskovsky, Artiom and Weissenberger, Felix and Rong, Keran and Tanno, Ryutaro and others},
  journal={arXiv preprint arXiv:2502.18864},
  year={2025}
}

@article{tang2025ai,
  title={AI-Researcher: Autonomous Scientific Innovation},
  author={Tang, Jiabin and Xia, Lianghao and Li, Zhonghang and Huang, Chao},
  journal={arXiv preprint arXiv:2505.18705},
  year={2025}
}

@article{mitchener2025kosmos,
  title={Kosmos: An ai scientist for autonomous discovery},
  author={Mitchener, Ludovico and Yiu, Angela and Chang, Benjamin and Bourdenx, Mathieu and Nadolski, Tyler and Sulovari, Arvis and Landsness, Eric C and Barabasi, Daniel L and Narayanan, Siddharth and Evans, Nicky and others},
  journal={arXiv preprint arXiv:2511.02824},
  year={2025}
}

@inproceedings{geva2021transformer,
  title={Transformer Feed-Forward Layers Are Key-Value Memories},
  url={http://dx.doi.org/10.18653/v1/2021.emnlp-main.446},
  doi={10.18653/v1/2021.emnlp-main.446},
  booktitle={Proceedings of the 2021 Conference on Empirical Methods in Natural Language Processing},
  publisher={Association for Computational Linguistics},
  author={Geva, Mor and Schuster, Roei and Berant, Jonathan and Levy, Omer},
  year={2021},
  pages={5484–5495}
}

@misc{weiss2021thinking,
  doi={10.48550/ARXIV.2106.06981},
  url={https://arxiv.org/abs/2106.06981},
  author={Weiss, Gail and Goldberg, Yoav and Yahav, Eran},
  title={Thinking Like Transformers},
  publisher={arXiv},
  year={2021}
}

@inproceedings{lindner2023tracr,
  series={NeurIPS 2023},
  title={Tracr: Compiled Transformers as a Laboratory for Interpretability},
  url={http://dx.doi.org/10.52202/075280-1649},
  doi={10.52202/075280-1649},
  booktitle={Advances in Neural Information Processing Systems 36},
  publisher={Neural Information Processing Systems Foundation, Inc. (NeurIPS)},
  author={Lindner, David and Kramar, Janos and Farquhar, Sebastian and Rahtz, Matthew and McGrath, Tom and Mikulik, Vladimir},
  year={2023},
  pages={37876–37899},
  collection={NeurIPS 2023}
}

@inproceedings{friedman2023learning,
  series={NeurIPS 2023},
  title={Learning Transformer Programs},
  url={http://dx.doi.org/10.52202/075280-2131},
  doi={10.52202/075280-2131},
  booktitle={Advances in Neural Information Processing Systems 36},
  publisher={Neural Information Processing Systems Foundation, Inc. (NeurIPS)},
  author={Friedman, Dan and Wettig, Alexander and Chen, Danqi},
  year={2023},
  pages={49044–49067},
  collection={NeurIPS 2023}
}

@misc{nanda2023progress,
  doi={10.48550/ARXIV.2301.05217},
  url={https://arxiv.org/abs/2301.05217},
  author={Nanda, Neel and Chan, Lawrence and Lieberum, Tom and Smith, Jess and Steinhardt, Jacob},
  title={Progress measures for grokking via mechanistic interpretability},
  publisher={arXiv},
  year={2023}
}

@misc{quirke2023understanding,
  doi={10.48550/ARXIV.2310.13121},
  url={https://arxiv.org/abs/2310.13121},
  author={Quirke, Philip and Barez, Fazl},
  title={Understanding Addition in Transformers},
  publisher={arXiv},
  year={2023}
}

@misc{takehi2026weightspace,
  title        = {{Sparks of Weight-Space Programming: LLMs can natively read \& write neural network weights}},
  author       = {Takehi, Rikiya and Teoh, Jayden and Zhang, Alex and Khattab, Omar},
  year         = {2026},
  month        = nov,
  howpublished = {\url{https://rikiyat.github.io/blog/weight-space-programming/}},
  note         = {Blog post}
}

@misc{team2026scaling,
  doi={10.48550/ARXIV.2608.07594},
  url={https://arxiv.org/abs/2608.07594},
  author={{Guide Labs Team} and Madsen, Andreas and Ismail, Aya Abdelsalam and Nguyen, Giang and Plant, Isaac and Chaudhary, Muawiz and Monson, Nathaniel and Azim, Saqib and Guo, Zhichen and Adebayo, Julius},
  title={Scaling Inherently Interpretable Language Models},
  publisher={arXiv},
  year={2026}
}

@article{tang2023semantic,
  title={Semantic reconstruction of continuous language from non-invasive brain recordings},
  volume={26},
  url={http://dx.doi.org/10.1038/s41593-023-01304-9},
  doi={10.1038/s41593-023-01304-9},
  number={5},
  journal={Nature Neuroscience},
  publisher={Springer Science and Business Media LLC},
  author={Tang, Jerry and LeBel, Amanda and Jain, Shailee and Huth, Alexander G.},
  year={2023},
  pages={858–866}
}

@inproceedings{antonello2023scaling,
  series={NeurIPS 2023},
  title={Scaling laws for language encoding models in fMRI},
  url={http://dx.doi.org/10.52202/075280-0958},
  doi={10.52202/075280-0958},
  booktitle={Advances in Neural Information Processing Systems 36},
  publisher={Neural Information Processing Systems Foundation, Inc. (NeurIPS)},
  author={Antonello, Richard and Vaidya, Aditya and Huth, Alexander},
  year={2023},
  pages={21895–21907},
  collection={NeurIPS 2023}
}

@article{jain2024computational,
  title={Computational Language Modeling and the Promise of In Silico Experimentation},
  volume={5},
  url={http://dx.doi.org/10.1162/nol_a_00101},
  doi={10.1162/nol_a_00101},
  number={1},
  journal={Neurobiology of Language},
  publisher={MIT Press},
  author={Jain, Shailee and Vo, Vy A. and Wehbe, Leila and Huth, Alexander G.},
  year={2024},
  pages={80–106}
}

@misc{georgiev2025mathematical,
  doi={10.48550/ARXIV.2511.02864},
  url={https://arxiv.org/abs/2511.02864},
  author={Georgiev, Bogdan and Gómez-Serrano, Javier and Tao, Terence and Wagner, Adam Zsolt},
  title={Mathematical exploration and discovery at scale},
  publisher={arXiv},
  year={2025}
}

@article{kasenberg2026ai,
  title={AI-Discovered Cognitive Models Reveal Novel Insights into Human and Animal Learning},
  url={http://dx.doi.org/10.64898/2026.05.18.725921},
  doi={10.64898/2026.05.18.725921},
  publisher={openRxiv},
  author={Kasenberg, Daniel and Castro, Pablo Samuel and Eckstein, Maria K. and Éltető, Nóemi and Dabney, Will and Wang, Caroline and Engelcke, Martin and Mohanta, Rishika and Dev, Aparna and Botvinick, Matthew M. and Tomasev, Nenad and Turner, Glenn C. and Costa, Vincent and Daw, Nathaniel D. and Stachenfeld, Kimberly L. and Miller, Kevin J.},
  year={2026}
}

@article{mcculloch1943logical,
  title={A logical calculus of the ideas immanent in nervous activity},
  volume={5},
  url={http://dx.doi.org/10.1007/BF02478259},
  doi={10.1007/bf02478259},
  number={4},
  journal={The Bulletin of Mathematical Biophysics},
  publisher={Springer Science and Business Media LLC},
  author={McCulloch, Warren S. and Pitts, Walter},
  year={1943},
  pages={115–133}
}

@misc{kim2024interpretable,
  doi={10.48550/ARXIV.2411.00066},
  url={https://arxiv.org/abs/2411.00066},
  author={Kim, Eunji and Mantena, Sriya and Yang, Weiwei and Singh, Chandan and Yoon, Sungroh and Gao, Jianfeng},
  title={Interpretable Next-token Prediction via the Generalized Induction Head},
  publisher={arXiv},
  year={2024}
}

@misc{singh2026agentic,
  doi={10.48550/ARXIV.2605.03808},
  url={https://arxiv.org/abs/2605.03808},
  author={Singh, Chandan and Tan, Yan Shuo and Xu, Weijia and Gero, Zelalem and Yang, Weiwei and Galley, Michel and Gao, Jianfeng},
  title={Agentic-imodels: Evolving agentic interpretability tools via autoresearch},
  publisher={arXiv},
  year={2026}
}

@article{rudin2019stop,
  title={Stop explaining black box machine learning models for high stakes decisions and use interpretable models instead},
  volume={1},
  url={http://dx.doi.org/10.1038/s42256-019-0048-x},
  doi={10.1038/s42256-019-0048-x},
  number={5},
  journal={Nature Machine Intelligence},
  publisher={Springer Science and Business Media LLC},
  author={Rudin, Cynthia},
  year={2019},
  pages={206–215}
}

@article{yao1975analysis,
  title={Analysis of the subtractive algorithm for greatest common divisors},
  volume={72},
  url={http://dx.doi.org/10.1073/pnas.72.12.4720},
  doi={10.1073/pnas.72.12.4720},
  number={12},
  journal={Proceedings of the National Academy of Sciences},
  publisher={National Academy of Sciences},
  author={Yao, Andrew C. and Knuth, Donald E.},
  year={1975},
  pages={4720–4722}
}

@article{ziems2024can,
  title={Can large language models transform computational social science?},
  author={Ziems, Caleb and Held, William and Shaikh, Omar and Chen, Jiaao and Zhang, Zhehao and Yang, Diyi},
  journal={Computational Linguistics},
  volume={50},
  number={1},
  pages={237--291},
  year={2024},
  publisher={MIT Press One Broadway, 12th Floor, Cambridge, Massachusetts 02142, USA~…}
}

@article{goodman2017european,
  title={European Union regulations on algorithmic decision-making and a “right to explanation”},
  author={Goodman, Bryce and Flaxman, Seth},
  journal={AI magazine},
  volume={38},
  number={3},
  pages={50--57},
  year={2017}
}

@article{singh2024rethinking,
  title={Rethinking interpretability in the era of large language models},
  author={Singh, Chandan and Inala, Jeevana Priya and Galley, Michel and Caruana, Rich and Gao, Jianfeng},
  journal={arXiv preprint arXiv:2402.01761},
  year={2024}
}

@article{caucheteux2023evidence,
  title={Evidence of a predictive coding hierarchy in the human brain listening to speech},
  volume={7},
  url={http://dx.doi.org/10.1038/s41562-022-01516-2},
  doi={10.1038/s41562-022-01516-2},
  number={3},
  journal={Nature Human Behaviour},
  publisher={Springer Science and Business Media LLC},
  author={Caucheteux, Charlotte and Gramfort, Alexandre and King, Jean-Rémi},
  year={2023},
  pages={430–441}
}

@article{goldstein2022shared,
  title={Shared computational principles for language processing in humans and deep language models},
  volume={25},
  url={http://dx.doi.org/10.1038/s41593-022-01026-4},
  doi={10.1038/s41593-022-01026-4},
  number={3},
  journal={Nature Neuroscience},
  publisher={Springer Science and Business Media LLC},
  author={Goldstein, Ariel and Zada, Zaid and Buchnik, Eliav and Schain, Mariano and Price, Amy and Aubrey, Bobbi and Nastase, Samuel A. and Feder, Amir and Emanuel, Dotan and Cohen, Alon and Jansen, Aren and Gazula, Harshvardhan and Choe, Gina and Rao, Aditi and Kim, Catherine and Casto, Colton and Fanda, Lora and Doyle, Werner and Friedman, Daniel and Dugan, Patricia and Melloni, Lucia and Reichart, Roi and Devore, Sasha and Flinker, Adeen and Hasenfratz, Liat and Levy, Omer and Hassidim, Avinatan and Brenner, Michael and Matias, Yossi and Norman, Kenneth A. and Devinsky, Orrin and Hasson, Uri},
  year={2022},
  pages={369–380}
}

@misc{toneva2019interpreting,
  doi={10.48550/ARXIV.1905.11833},
  url={https://arxiv.org/abs/1905.11833},
  author={Toneva, Mariya and Wehbe, Leila},
  title={Interpreting and improving natural-language processing (in machines) with natural language-processing (in the brain)},
  publisher={arXiv},
  year={2019}
}

@misc{sun2024concept,
  doi={10.48550/ARXIV.2412.07992},
  url={https://arxiv.org/abs/2412.07992},
  author={Sun, Chung-En and Oikarinen, Tuomas and Ustun, Berk and Weng, Tsui-Wei},
  title={Concept Bottleneck Large Language Models},
  publisher={arXiv},
  year={2024}
}

@misc{radford2022robust,
  doi={10.48550/ARXIV.2212.04356},
  url={https://arxiv.org/abs/2212.04356},
  author={Radford, Alec and Kim, Jong Wook and Xu, Tao and Brockman, Greg and McLeavey, Christine and Sutskever, Ilya},
  title={Robust Speech Recognition via Large-Scale Weak Supervision},
  publisher={arXiv},
  year={2022}
}

@article{chi2005multiresolution,
  title={Multiresolution spectrotemporal analysis of complex sounds},
  volume={118},
  url={http://dx.doi.org/10.1121/1.1945807},
  doi={10.1121/1.1945807},
  number={2},
  journal={The Journal of the Acoustical Society of America},
  publisher={Acoustical Society of America (ASA)},
  author={Chi, Taishih and Ru, Powen and Shamma, Shihab A.},
  year={2005},
  pages={887–906}
}

@article{anden2013deep,
  doi={10.48550/ARXIV.1304.6763},
  url={https://arxiv.org/abs/1304.6763},
  author={Andén, Joakim and Mallat, Stéphane},
  title={Deep Scattering Spectrum},
  publisher={arXiv},
  year={2013}
}

@misc{ha2016hypernetworks,
  doi={10.48550/ARXIV.1609.09106},
  url={https://arxiv.org/abs/1609.09106},
  author={Ha, David and Dai, Andrew and Le, Quoc V.},
  title={HyperNetworks},
  publisher={arXiv},
  year={2016}
}

@article{stanley2009hypercube,
  title={A Hypercube-Based Encoding for Evolving Large-Scale Neural Networks},
  volume={15},
  url={http://dx.doi.org/10.1162/artl.2009.15.2.15202},
  doi={10.1162/artl.2009.15.2.15202},
  number={2},
  journal={Artificial Life},
  publisher={MIT Press - Journals},
  author={Stanley, Kenneth O. and D’Ambrosio, David B. and Gauci, Jason},
  year={2009},
  pages={185–212}
}

@article{chauhan2024brief,
  title={A brief review of hypernetworks in deep learning},
  volume={57},
  url={http://dx.doi.org/10.1007/s10462-024-10862-8},
  doi={10.1007/s10462-024-10862-8},
  number={9},
  journal={Artificial Intelligence Review},
  publisher={Springer Science and Business Media LLC},
  author={Chauhan, Vinod Kumar and Zhou, Jiandong and Lu, Ping and Molaei, Soheila and Clifton, David A.},
  year={2024}
}

@inproceedings{schurholt2022hyper,
  series={NeurIPS 2022},
  title={Hyper-Representations as Generative Models: Sampling Unseen Neural Network Weights},
  url={http://dx.doi.org/10.52202/068431-2024},
  doi={10.52202/068431-2024},
  booktitle={Advances in Neural Information Processing Systems 35},
  publisher={Neural Information Processing Systems Foundation, Inc. (NeurIPS)},
  author={Schürholt, Konstantin and Knyazev, Boris and Giró-I-Nieto, Xavier and Borth, Damian},
  year={2022},
  pages={27906–27920},
  collection={NeurIPS 2022}
}

@misc{peebles2022learning,
  doi={10.48550/ARXIV.2209.12892},
  url={https://arxiv.org/abs/2209.12892},
  author={Peebles, William and Radosavovic, Ilija and Brooks, Tim and Efros, Alexei A. and Malik, Jitendra},
  title={Learning to Learn with Generative Models of Neural Network Checkpoints},
  publisher={arXiv},
  year={2022}
}

@misc{knyazev2021parameter,
  doi={10.48550/ARXIV.2110.13100},
  url={https://arxiv.org/abs/2110.13100},
  author={Knyazev, Boris and Drozdzal, Michal and Taylor, Graham W. and Romero-Soriano, Adriana},
  title={Parameter Prediction for Unseen Deep Architectures},
  publisher={arXiv},
  year={2021}
}

@misc{navon2023equivariant,
  doi={10.48550/ARXIV.2301.12780},
  url={https://arxiv.org/abs/2301.12780},
  author={Navon, Aviv and Shamsian, Aviv and Achituve, Idan and Fetaya, Ethan and Chechik, Gal and Maron, Haggai},
  title={Equivariant Architectures for Learning in Deep Weight Spaces},
  publisher={arXiv},
  year={2023}
}

@misc{han2026survey,
  doi={10.48550/ARXIV.2603.10090},
  url={https://arxiv.org/abs/2603.10090},
  author={Han, Xiaolong and Wang, Zehong and Zhao, Bo and Zhang, Binchi and Li, Jundong and Borth, Damian and Yu, Rose and Maron, Haggai and Ye, Yanfang and Yin, Lu and Neri, Ferrante},
  title={A Survey of Weight Space Learning: Understanding, Representation, and Generation},
  publisher={arXiv},
  year={2026}
}

@misc{paulo2024automatically,
  doi={10.48550/ARXIV.2410.13928},
  url={https://arxiv.org/abs/2410.13928},
  author={Paulo, Gonçalo and Mallen, Alex and Juang, Caden and Belrose, Nora},
  title={Automatically Interpreting Millions of Features in Large Language Models},
  publisher={arXiv},
  year={2024}
}

@misc{liu2026neuronscope,
  doi={10.48550/ARXIV.2601.03671},
  url={https://arxiv.org/abs/2601.03671},
  author={Liu, Weiqi and Miao, Yongliang and Zhao, Haiyan and Liu, Yanguang and Du, Mengnan},
  title={NeuronScope: A Multi-Agent Framework for Explaining Polysemantic Neurons in Language Models},
  publisher={arXiv},
  year={2026}
}

@misc{marinllobet2026automated,
  doi={10.48550/ARXIV.2605.01555},
  url={https://arxiv.org/abs/2605.01555},
  author={Marin-Llobet, Arnau and Ferrando, Javier},
  title={Automated Interpretability and Feature Discovery in Language Models with Agents},
  publisher={arXiv},
  year={2026}
}

@misc{zaigrajew2026line,
  doi={10.48550/ARXIV.2604.08039},
  url={https://arxiv.org/abs/2604.08039},
  author={Zaigrajew, Vladimir and Piechota, Michał and Sekula, Gaspar and Gelar, Paweł and Biecek, Przemysław},
  title={LINE: LLM-based Iterative Neuron Explanations for Vision Models},
  publisher={arXiv},
  year={2026}
}

@misc{khan2026can,
  doi={10.48550/ARXIV.2606.24026},
  url={https://arxiv.org/abs/2606.24026},
  author={Khan, Ayan Antik and Kohli, Harsh and Yao, Yuekun and Sun, Huan and Yao, Ziyu},
  title={Can Language Model Agents be Helpful Circuit Explainers in Mechanistic Interpretability?},
  publisher={arXiv},
  year={2026}
}

@inproceedings{socher2013recursive,
  title={Recursive Deep Models for Semantic Compositionality Over a Sentiment Treebank},
  url={http://dx.doi.org/10.18653/v1/d13-1170},
  doi={10.18653/v1/d13-1170},
  booktitle={Proceedings of the 2013 Conference on Empirical Methods in Natural Language Processing},
  publisher={Association for Computational Linguistics},
  author={Socher, Richard and Perelygin, Alex and Wu, Jean and Chuang, Jason and Manning, Christopher D. and Ng, Andrew and Potts, Christopher},
  year={2013},
  pages={1631–1642}
}

@inproceedings{li2002learning,
  title={Learning question classifiers},
  volume={1},
  url={http://dx.doi.org/10.3115/1072228.1072378},
  doi={10.3115/1072228.1072378},
  booktitle={Proceedings of the 19th international conference on Computational linguistics  -},
  publisher={Association for Computational Linguistics},
  author={Li, Xin and Roth, Dan},
  year={2002},
  pages={1–7}
}

@inproceedings{dolan2004unsupervised,
  series={COLING ’04},
  title={Unsupervised construction of large paraphrase corpora: exploiting massively parallel news sources},
  url={http://dx.doi.org/10.3115/1220355.1220406},
  doi={10.3115/1220355.1220406},
  booktitle={Proceedings of the 20th international conference on Computational Linguistics  - COLING ’04},
  publisher={Association for Computational Linguistics},
  author={Dolan, Bill and Quirk, Chris and Brockett, Chris},
  year={2004},
  pages={350–es},
  collection={COLING ’04}
}

@inproceedings{bowman2015large,
  title={A large annotated corpus for learning natural language inference},
  url={http://dx.doi.org/10.18653/v1/D15-1075},
  doi={10.18653/v1/d15-1075},
  booktitle={Proceedings of the 2015 Conference on Empirical Methods in Natural Language Processing},
  publisher={Association for Computational Linguistics},
  author={Bowman, Samuel R. and Angeli, Gabor and Potts, Christopher and Manning, Christopher D.},
  year={2015},
  pages={632–642}
}

@misc{moussa2024improving,
  doi={10.48550/ARXIV.2410.09230},
  url={https://arxiv.org/abs/2410.09230},
  author={Moussa, Omer and Klakow, Dietrich and Toneva, Mariya},
  title={Improving Semantic Understanding in Speech Language Models via Brain-tuning},
  publisher={arXiv},
  year={2024}
}

@article{zada2025podcast,
  title={The “Podcast” ECoG dataset for modeling neural activity during natural language comprehension},
  url={http://dx.doi.org/10.1101/2025.02.14.638352},
  doi={10.1101/2025.02.14.638352},
  publisher={openRxiv},
  author={Zada, Zaid and Nastase, Samuel A. and Aubrey, Bobbi and Jalon, Itamar and Michelmann, Sebastian and Wang, Haocheng and Hasenfratz, Liat and Doyle, Werner and Friedman, Daniel and Dugan, Patricia and Melloni, Lucia and Devore, Sasha and Flinker, Adeen and Devinsky, Orrin and Goldstein, Ariel and Hasson, Uri},
  year={2025}
}

@article{abnar2019,
  author       = {Samira Abnar and
                  Lisa Beinborn and
                  Rochelle Choenni and
                  Willem H. Zuidema},
  title        = {Blackbox meets blackbox: Representational Similarity and Stability
                  Analysis of Neural Language Models and Brains},
  journal      = {CoRR},
  volume       = {abs/1906.01539},
  year         = {2019},
  url          = {http://arxiv.org/abs/1906.01539},
  eprinttype    = {arXiv},
  eprint       = {1906.01539},
  bibsource    = {dblp computer science bibliography, https://dblp.org}
}

@article{kamps2016occipital,
  title={The occipital place area represents the local elements of scenes},
  author={Kamps, Frederik S and Julian, Joshua B and Kubilius, Jonas and Kanwisher, Nancy and Dilks, Daniel D},
  journal={Neuroimage},
  volume={132},
  pages={417--424},
  year={2016},
  publisher={Elsevier}
}

@article{kanwisher2006fusiform,
  title={The fusiform face area: a cortical region specialized for the perception of faces},
  author={Kanwisher, Nancy and Yovel, Galit},
  journal={Philosophical Transactions of the Royal Society B: Biological Sciences},
  volume={361},
  number={1476},
  pages={2109--2128},
  year={2006},
  publisher={The Royal Society London}
}

@misc{bojanowski2016enriching,
  doi={10.48550/ARXIV.1607.04606},
  url={https://arxiv.org/abs/1607.04606},
  author={Bojanowski, Piotr and Grave, Edouard and Joulin, Armand and Mikolov, Tomas},
  title={Enriching Word Vectors with Subword Information},
  publisher={arXiv},
  year={2016}
}

@misc{ludan2023interpretable,
  doi={10.48550/ARXIV.2310.19660},
  url={https://arxiv.org/abs/2310.19660},
  author={Ludan, Josh Magnus and Lyu, Qing and Yang, Yue and Dugan, Liam and Yatskar, Mark and Callison-Burch, Chris},
  title={Interpretable-by-Design Text Understanding with Iteratively Generated Concept Bottleneck},
  publisher={arXiv},
  year={2023}
}

@inbook{tan2024interpreting,
  title={Interpreting Pretrained Language Models via Concept Bottlenecks},
  url={http://dx.doi.org/10.1007/978-981-97-2259-4_5},
  doi={10.1007/978-981-97-2259-4_5},
  booktitle={Advances in Knowledge Discovery and Data Mining},
  publisher={Springer Nature Singapore},
  author={Tan, Zhen and Cheng, Lu and Wang, Song and Yuan, Bo and Li, Jundong and Liu, Huan},
  year={2024},
  pages={56–74}
}

@misc{gao2024scaling,
  doi={10.48550/ARXIV.2406.04093},
  url={https://arxiv.org/abs/2406.04093},
  author={Gao, Leo and la Tour, Tom Dupré and Tillman, Henk and Goh, Gabriel and Troll, Rajan and Radford, Alec and Sutskever, Ilya and Leike, Jan and Wu, Jeffrey},
  title={Scaling and evaluating sparse autoencoders},
  publisher={arXiv},
  year={2024}
}

@misc{turner2023steering,
  doi={10.48550/ARXIV.2308.10248},
  url={https://arxiv.org/abs/2308.10248},
  author={Turner, Alexander Matt and Thiergart, Lisa and Leech, Gavin and Udell, David and Vazquez, Juan J. and Mini, Ulisse and MacDiarmid, Monte},
  title={Steering Language Models With Activation Engineering},
  publisher={arXiv},
  year={2023}
}

@misc{li2023inference,
  doi={10.48550/ARXIV.2306.03341},
  url={https://arxiv.org/abs/2306.03341},
  author={Li, Kenneth and Patel, Oam and Viégas, Fernanda and Pfister, Hanspeter and Wattenberg, Martin},
  title={Inference-Time Intervention: Eliciting Truthful Answers from a Language Model},
  publisher={arXiv},
  year={2023}
}

@article{mitchell2008predicting,
  title={Predicting Human Brain Activity Associated with the Meanings of Nouns},
  volume={320},
  url={http://dx.doi.org/10.1126/science.1152876},
  doi={10.1126/science.1152876},
  number={5880},
  journal={Science},
  publisher={American Association for the Advancement of Science (AAAS)},
  author={Mitchell, Tom M. and Shinkareva, Svetlana V. and Carlson, Andrew and Chang, Kai-Min and Malave, Vicente L. and Mason, Robert A. and Just, Marcel Adam},
  year={2008},
  pages={1191–1195}
}

@article{wehbe2014simultaneously,
  title={Simultaneously Uncovering the Patterns of Brain Regions Involved in Different Story Reading Subprocesses},
  volume={9},
  url={http://dx.doi.org/10.1371/journal.pone.0112575},
  doi={10.1371/journal.pone.0112575},
  number={11},
  journal={PLoS ONE},
  publisher={Public Library of Science (PLoS)},
  author={Wehbe, Leila and Murphy, Brian and Talukdar, Partha and Fyshe, Alona and Ramdas, Aaditya and Mitchell, Tom},
  editor={Paterson, Kevin},
  year={2014},
  pages={e112575}
}

@article{nishida2018decoding,
  title={Decoding naturalistic experiences from human brain activity via distributed representations of words},
  volume={180},
  url={http://dx.doi.org/10.1016/j.neuroimage.2017.08.017},
  doi={10.1016/j.neuroimage.2017.08.017},
  journal={NeuroImage},
  publisher={Elsevier BV},
  author={Nishida, Satoshi and Nishimoto, Shinji},
  year={2018},
  pages={232–242}
}

@article{kumar2022shared,
  title={Shared functional specialization in transformer-based language models and the human brain},
  url={http://dx.doi.org/10.1101/2022.06.08.495348},
  doi={10.1101/2022.06.08.495348},
  publisher={openRxiv},
  author={Kumar, Sreejan and Sumers, Theodore R. and Yamakoshi, Takateru and Goldstein, Ariel and Hasson, Uri and Norman, Kenneth A. and Griffiths, Thomas L. and Hawkins, Robert D. and Nastase, Samuel A.},
  year={2022}
}

@article{thurnherr2024neural,
  doi={10.48550/ARXIV.2410.00061},
  url={https://arxiv.org/abs/2410.00061},
  author={Thurnherr, Hannes and Riesen, Kaspar},
  title={Neural Decompiling of Tracr Transformers},
  publisher={arXiv},
  year={2024}
}

@misc{huang2026discovering,
  doi={10.48550/ARXIV.2602.08857},
  url={https://arxiv.org/abs/2602.08857},
  author={Huang, Xinting and Bakalova, Aleksandra and Bhattamishra, Satwik and Merrill, William and Hahn, Michael},
  title={Discovering Interpretable Algorithms by Decompiling Transformers to RASP},
  publisher={arXiv},
  year={2026}
}

@misc{skalse2022defining,
  doi={10.48550/ARXIV.2209.13085},
  url={https://arxiv.org/abs/2209.13085},
  author={Skalse, Joar and Howe, Nikolaus H. R. and Krasheninnikov, Dmitrii and Krueger, David},
  title={Defining and Characterizing Reward Hacking},
  publisher={arXiv},
  year={2022}
}

@misc{bergen2026monitoring,
  doi={10.48550/ARXIV.2609.19101},
  url={https://arxiv.org/abs/2609.19101},
  author={Bergen, Leon and Bhalla, Usha and Lee, Andrew and Widawsky, Barak and Nasvytis, Linas and Watts, Connor and Boppana, Siddharth and Baskaran, Sidharth and Hazra, Dron and Byun, Michael and Geiger, Atticus and Lewis, Owen and Kowal, Matthew and Shyam, Vasudev and Fel, Thomas and McGrath, Thomas and Lubana, Ekdeep Singh and Merullo, Jack},
  title={Monitoring and Discovering Reward Hacking with Internal Representations during LLM Evaluations},
  publisher={arXiv},
  year={2026}
}

@misc{elhage2022toy,
  doi={10.48550/ARXIV.2209.10652},
  url={https://arxiv.org/abs/2209.10652},
  author={Elhage, Nelson and Hume, Tristan and Olsson, Catherine and Schiefer, Nicholas and Henighan, Tom and Kravec, Shauna and Hatfield-Dodds, Zac and Lasenby, Robert and Drain, Dawn and Chen, Carol and Grosse, Roger and McCandlish, Sam and Kaplan, Jared and Amodei, Dario and Wattenberg, Martin and Olah, Christopher},
  title={Toy Models of Superposition},
  publisher={arXiv},
  year={2022}
}

@misc{chughtai2023toy,
  doi={10.48550/ARXIV.2302.03025},
  url={https://arxiv.org/abs/2302.03025},
  author={Chughtai, Bilal and Chan, Lawrence and Nanda, Neel},
  title={A Toy Model of Universality: Reverse Engineering How Networks Learn Group Operations},
  publisher={arXiv},
  year={2023}
}

@inproceedings{teso2019explanatory,
  series={AIES ’19},
  title={Explanatory Interactive Machine Learning},
  url={http://dx.doi.org/10.1145/3306618.3314293},
  doi={10.1145/3306618.3314293},
  booktitle={Proceedings of the 2019 AAAI/ACM Conference on AI, Ethics, and Society},
  publisher={ACM},
  author={Teso, Stefano and Kersting, Kristian},
  year={2019},
  pages={239–245},
  collection={AIES ’19}
}

@article{amershi2014power,
  title={Power to the People: The Role of Humans in Interactive Machine Learning},
  volume={35},
  url={http://dx.doi.org/10.1609/aimag.v35i4.2513},
  doi={10.1609/aimag.v35i4.2513},
  number={4},
  journal={AI Magazine},
  publisher={Wiley},
  author={Amershi, Saleema and Cakmak, Maya and Knox, W. Bradley and Kulesza, Todd},
  year={2014},
  pages={105–120}
}

@article{holzinger2016interactive,
  title={Interactive machine learning for health informatics: when do we need the human-in-the-loop?},
  volume={3},
  url={http://dx.doi.org/10.1007/s40708-016-0042-6},
  doi={10.1007/s40708-016-0042-6},
  number={2},
  journal={Brain Informatics},
  publisher={Springer Science and Business Media LLC},
  author={Holzinger, Andreas},
  year={2016},
  pages={119–131}
}

@inbook{lewis1998naive,
  title={Naive (Bayes) at forty: The independence assumption in information retrieval},
  url={http://dx.doi.org/10.1007/bfb0026666},
  doi={10.1007/bfb0026666},
  booktitle={Machine Learning: ECML-98},
  publisher={Springer Berlin Heidelberg},
  author={Lewis, David D.},
  year={1998},
  pages={4–15}
}

@misc{papailiopoulos2026adderboard,
  author       = {Papailiopoulos, Dimitris},
  title        = {{AdderBoard}: A Community Challenge for Tiny Arithmetic Transformers},
  year         = {2026},
  publisher    = {GitHub},
  howpublished = {\url{https://github.com/anadim/AdderBoard}},
  note         = {Open challenge and leaderboard for building the smallest transformers that can perform integer addition}
}

@misc{binhuraib2025litcoder,
  doi={10.48550/ARXIV.2509.09152},
  url={https://arxiv.org/abs/2509.09152},
  author={Binhuraib, Taha and Gao, Ruimin and Ivanova, Anna A.},
  title={LITcoder: A General-Purpose Library for Building and Comparing Encoding Models},
  publisher={arXiv},
  year={2025}
}

@article{portilla2000parametric,
  title={A Parametric Texture Model Based on Joint Statistics of Complex Wavelet Coefficients},
  volume={40},
  url={http://dx.doi.org/10.1023/a:1026553619983},
  doi={10.1023/a:1026553619983},
  number={1},
  journal={International Journal of Computer Vision},
  publisher={Springer Science and Business Media LLC},
  author={Portilla, Javier and Simoncelli, Eero P.},
  year={2000},
  pages={49–70}
}

@article{oliva2001modeling,
  title={Modeling the Shape of the Scene: A Holistic Representation of the Spatial Envelope},
  volume={42},
  url={http://dx.doi.org/10.1023/a:1011139631724},
  doi={10.1023/a:1011139631724},
  number={3},
  journal={International Journal of Computer Vision},
  publisher={Springer Science and Business Media LLC},
  author={Oliva, Aude and Torralba, Antonio},
  year={2001},
  pages={145–175}
}

@article{kay2008identifying,
  title={Identifying natural images from human brain activity},
  volume={452},
  url={http://dx.doi.org/10.1038/nature06713},
  doi={10.1038/nature06713},
  number={7185},
  journal={Nature},
  publisher={Springer Science and Business Media LLC},
  author={Kay, Kendrick N. and Naselaris, Thomas and Prenger, Ryan J. and Gallant, Jack L.},
  year={2008},
  pages={352–355}
}

@article{heeger1992normalization,
  title={Normalization of cell responses in cat striate cortex},
  volume={9},
  url={http://dx.doi.org/10.1017/s0952523800009640},
  doi={10.1017/s0952523800009640},
  number={2},
  journal={Visual Neuroscience},
  publisher={Maximum Academic Press},
  author={Heeger, David J.},
  year={1992},
  pages={181–197}
}

@misc{zerbe2026laion,
  author       = {Zerbe, J. and Roth, J. and Mell, M. M. and Herholz, P. and Knapen, T. and Hebart, M. N.},
  title        = {{LAION-fMRI}: A densely sampled {7T-fMRI} dataset providing broad coverage of natural image diversity},
  howpublished = {Talk 25.11, Scene Perception Talk Session, Vision Sciences Society Annual Meeting},
  month        = may,
  year         = {2026},
  note         = {VSS abstract}
}
